\documentclass[12pt]{iopart}
\usepackage{graphicx}

\begin{document}

\title[Branching fraction, $CP$ asymmetry, and polarization in $B\to\rho\rho$ decays]{Branching fraction, $CP$ asymmetry, and polarization in $B\to\rho\rho$ decays with the modified perturbative QCD approach}

\author{Ru-Xuan Wang $^1$\footnote{Author to whom any correspondence should be addressed}, 
	Mao-Zhi Yang $^1$ \footnote{Author to whom any correspondence should be addressed}}
\address{$^1$ School of Physics, Nankai University, \\Tianjin 300071, China}
\eads{\mailto{wangrx@mail.nankai.edu.cn}, \mailto{yangmz@nankai.edu.cn}}

\begin{abstract}
In this work, we examine the observables associated with the $B\to\rho\rho$ decays, including branching fractions, $CP$ asymmetry parameters, and longitudinal polarization fractions in the perturbative QCD (PQCD) approach with a few improvements. 
The essential distinction between this study and previous works lies in the introduction of an infrared cutoff at the critical scale $\mu_c$, which is approximately at the scale of $1\;\mathrm{GeV}$. 
The contributions above the critical scale are calculated using the PQCD approach, consistent with earlier studies, while the contributions below the scale $\mu_c$ are regarded as nonperturbative and represented by some soft form factors. 
In addition, the distribution amplitude of the $B$ meson, derived from the relativistic potential model, and the contributions from the color-octet quark-antiquark components are also taken into account. 
With these modifications, we find that the theoretical results agree well with the experimental measurements for most observables, the introduction of the infrared cutoff enhances the reliability of the perturbation calculations, and the color-octet contributions play a key role in explaining the experimental data.
\end{abstract}


\submitto{\jpg}
\maketitle

\section{Introduction \label{sec:intro}}
The investigation of $B$ meson decay improves our understanding of the Standard Model (SM) and offers opportunities to explore new physics beyond the SM. 
The weak decay of the $B$ meson is a complex problem, as it involves not only the weak interaction but also the strong interaction. 
In light of the complication of this issue, a variety of theoretical methods have been formulated, 
including the QCD factorization approach (QCDF) \cite{Beneke:1999br, Beneke:2000ry, Beneke:2001ev, Beneke:2003zv}, 
soft collinear effective theory (SCET) \cite{Bauer:2000yr, Bauer:2001cu, Bauer:2001yt}, and the perturbative QCD approach (PQCD) \cite{Keum:2000ph, Lu:2000em, Keum:2000wi}. 
With the operation and upgrading of the $B$ factory \cite{Belle-II:2018jsg, Belle-II:2022cgf} and LHCb experiment \cite{LHCb:2018roe, Cerri:2018ypt}, 
a significant amount of experimental data has been made available, refining most measurements of the $B$ meson decays and allowing us to examine these theoretical methods with higher accuracy and precision. 

To narrow the discrepancy between the experimental measurements and the theoretical calculations in $B\to\pi\pi$ and $K\pi$ decay channels, commonly referred to as the $\pi\pi$ and $K\pi$ puzzles \cite{Li:2005kt, Li:2009wba},  
a modification of the PQCD approach has been proposed recently by implementing an infrared cutoff at the scale of approximately 1 GeV \cite{Lu:2021gwt}. This approach has been demonstrated to be helpful in resolving the long-standing puzzles \cite{Lu:2022hbp, Wang:2022ihx}, 
and it additionally serves to interpret data from other $B\to PP$ and $PV$ decay modes \cite{Lu:2024jzn, Gui:2024nlk, Wang:2024xci}, where $P(V)$ denotes the light pseudoscalar (vector) meson. Given the close connection of $B\to\pi\pi$ and $\rho\rho$ decays, it is necessary to discuss whether this approach is capable of explaining the measurements of the $B\to\rho\rho$ decay. The $\rho\rho$ decay channel is of particular interest, which has been extensively studied in previous works \cite{Kagan:2004uw, Li:2004ti, Li:2005hg, Li:2006cva, Beneke:2006hg, Ali:2007ff, Zou:2015iwa, Wang:2017hxe, Yan:2018fif, Chai:2022ptk}. It has been pointed out in Ref. \cite{Li:2006cva} that the experimental data of $B\to \rho\rho$ have seriously constrained the resolution of $B\to\pi\pi$ puzzle in the theoretical approaches such as PQCD, QCD and SCET. On one hand, the next-to-leading order (NLO) PQCD predictions to the branching ratios $B\to \rho\rho$ decays agree with experimental data, while the predicted branching ratio of $B^0\to\pi^0\pi^0$ is several times smaller than experimental data \cite{Li:2006cva}. On the other hand, the QCDF approach with the input of SCET jet function predicts compatible value for the branching ratio of $B^0\to \pi^0\pi^0$ decay, while the prediction to the branching ratio of $B^0\to \rho^0\rho^0$ with the same method overshoots the experimental measurement very much \cite{Li:2006cva}. The modified PQCD approach can solve the $\pi\pi$, $K\pi$ puzzles and explain all the data of $B\to PP$ decays at the same time. Therefore, it is necessary to extend this approach to $B\to\rho\rho$ decay.

Here, based on the conventional PQCD calculations, we consider an infrared cutoff at the critical scale $\mu_c$ in this paper. 
The components with the scales exceeding the scale of $\mu_c$ are computed in PQCD approach, while those contributions that have the scales below the critical scale are regarded as of nonperturbative dynamics and are represented by some soft form factors.
What's more, the contributions with the quark-antiquark pairs in color-octet states in final state are also crucial to achieve a more precise alignment between the theoretical calculations and the experimental measurements. 
With the appropriate numerical parameters, the relevant observables of the $B\to\rho\rho$ channel,  including branching fractions, direct $CP$ asymmetries, 
and longitudinal polarization fractions, are in agreement with the experimental data within the bounds of uncertainty. 
It is noteworthy that the nonperturbative parameters employed here are similar to those used in the $\pi\pi$ and $K\pi$ cases. 
Although the underlying nature of the fitted parameters, i.e., how to calculate them under the first principle of QCD, requires further investigation, 
there is the possibility to account for all the charmless two-body decays of the $B$ meson within this framework by incorporating the SU(3) symmetry and its breaking effects \cite{Lu:2024jzn}. 

This paper is organized in the following manner.
In section~\ref{sec:lo}, we briefly explain the PQCD approach used in the calculation and present the expressions for the relevant Feynman diagrams at leading order. 
The various corrections arising from the NLO diagrams, soft form factors, and color-octet operators are introduced in section~\ref{sec:cor}. 
In section~\ref{sec:num}, we detail the input parameters employed in the numerical calculations, 
followed by the resultant branching fractions, direct $CP$ symmetries, and longitudinal polarization fractions. 
Furthermore, we compare the theoretical calculations with the experimental data and examine the influences from different sources in section~\ref{sec:num}. 
A concise summary is provided in section~\ref{sec:sum}. 

\section{Leading order calculation \label{sec:lo}}
\subsection{Theoretical framework \label{subsec:frame}}
To obtain the physical observables we are interested in, we need to calculate the decay amplitude of $B\to\rho\rho$ with the help of the effective Hamiltonian
\begin{eqnarray}
    \mathcal{M} \propto \langle \rho\rho | H_{\textrm{eff}} | B \rangle.
\end{eqnarray}
The decays of $B$ meson are governed by the weak decay of $b$ quark. 
Here, in $B\to\rho\rho$ decay, the $b\to d$ transition is concerned and the effective Hamiltonian is given by \cite{Buchalla:1995vs}
\begin{eqnarray}\label{eq:heff}
	\fl &\mathcal{H}_{\textrm{eff}}(b\to d)  \nonumber \\
	\fl & =\frac{G_F}{\sqrt{2}}\biggl\{V_u\left[C_1(\mu)O_1^u(\mu) + C_2(\mu)O_2^u(\mu)\right] 
		-V_t\sum_{i=3}^{10}\left[C_i(\mu)O_i(\mu)+C_{8g}(\mu)O_{8g}(\mu)\right]\biggr\}, 
\end{eqnarray}
where $G_F$ is the Fermi constant, $V_{u(t)}$ are combinations of the Cabibbo-Kobayashi-Maskawa (CKM) matrix elements with $V_{u(t)} = V_{{u(t)}b}V_{{u(t)}d}^*$. 
The Wilson coefficients $C_i$ and the four-quark operators $O_i$ are renormalization scale $\mu$ dependent. 
Their dependence is canceled each other, leaving the amplitude physical and $\mu$ independent. 
The operators $O_i$ involved in the $b\to d$ transition can be categorized into tree-level ($O_{1-2}^u$), QCD penguin-level ($O_{3-6}$), electroweak penguin-level ($O_{7-10}$), and the chromomagnetic operators ($O_{8g}$), which are listed below
\begin{eqnarray} \label{eq:o1-10}
	\fl O_1^u &= \left(\bar{d}_{\alpha}u_{\beta} \right)_{V-A}\left(\bar{u}_{\beta}b_{\alpha}\right)_{V-A}, \quad 
	&O_2^u= \left(\bar{d}_{\alpha}u_{\alpha}\right)_{V-A}\left(\bar{u}_{\beta}b_{\beta} \right)_{V-A}, \nonumber \\
	\fl O_3   &= (\bar{d}_{\alpha}b_{\alpha})_{V-A}\sum_{q^\prime}\left(\bar{q^\prime}_{\beta}q^\prime_{\beta} \right)_{V-A}, \quad
	&O_4  = (\bar{d}_{\alpha}b_{\beta})_{V-A}\sum_{q^\prime} \left(\bar{q^\prime}_{\beta}q^\prime_{\alpha}\right)_{V-A}, \nonumber \\
	\fl O_5   &= (\bar{d}_{\alpha}b_{\alpha})_{V-A}\sum_{q^\prime}\left(\bar{q^\prime}_{\beta}q^\prime_{\beta} \right)_{V+A}, \quad
	&O_6  = (\bar{d}_{\alpha}b_{\beta})_{V-A}\sum_{q^\prime} \left(\bar{q^\prime}_{\beta}q^\prime_{\alpha}\right)_{V+A}, \nonumber \\
	\fl O_7   &= \frac{3}{2}(\bar{d}_{\alpha}b_{\alpha})_{V-A}\sum_{q^\prime}e_{q^\prime}\left(\bar{q^\prime}_{\beta}q^\prime_{\beta} \right)_{V+A}, \quad
	&O_8  = \frac{3}{2}(\bar{d}_{\alpha}b_{\beta})_{V-A}\sum_{q^\prime}e_{q^\prime} \left(\bar{q^\prime}_{\beta}q^\prime_{\alpha}\right)_{V+A}, \nonumber \\
	\fl O_9   &= \frac{3}{2}(\bar{d}_{\alpha}b_{\alpha})_{V-A}\sum_{q^\prime}e_{q^\prime}\left(\bar{q^\prime}_{\beta}q^\prime_{\beta} \right)_{V-A}, \quad
	&O_{10}= \frac{3}{2}(\bar{d}_{\alpha}b_{\beta})_{V-A}\sum_{q^\prime}e_{q^\prime} \left(\bar{q^\prime}_{\beta}q^\prime_{\alpha}\right)_{V-A}, \nonumber \\
	\fl O_{8g}&=\frac{g_s}{8\pi^2}m_b\bar{d}_\alpha\sigma^{\mu\nu}(1+\gamma_5)T^a_{\alpha\beta}G^a_{\mu\nu}b_\beta,
\end{eqnarray}
with the color SU(3) generator $T^a_{\alpha\beta}$, and the sum over the repeated color indices $\alpha$ and $\beta$ implied.
The subscript $V\pm A$ denote the chiral current structures $\gamma_\mu(1\pm \gamma_5)$ conventionally.
The sum of $q^\prime$ includes all the active quark flavors, i.e. $q^\prime \in \{u,d,s,c,b\}$.

\begin{figure}[t]
	\raggedleft
	\includegraphics[width=\textwidth]{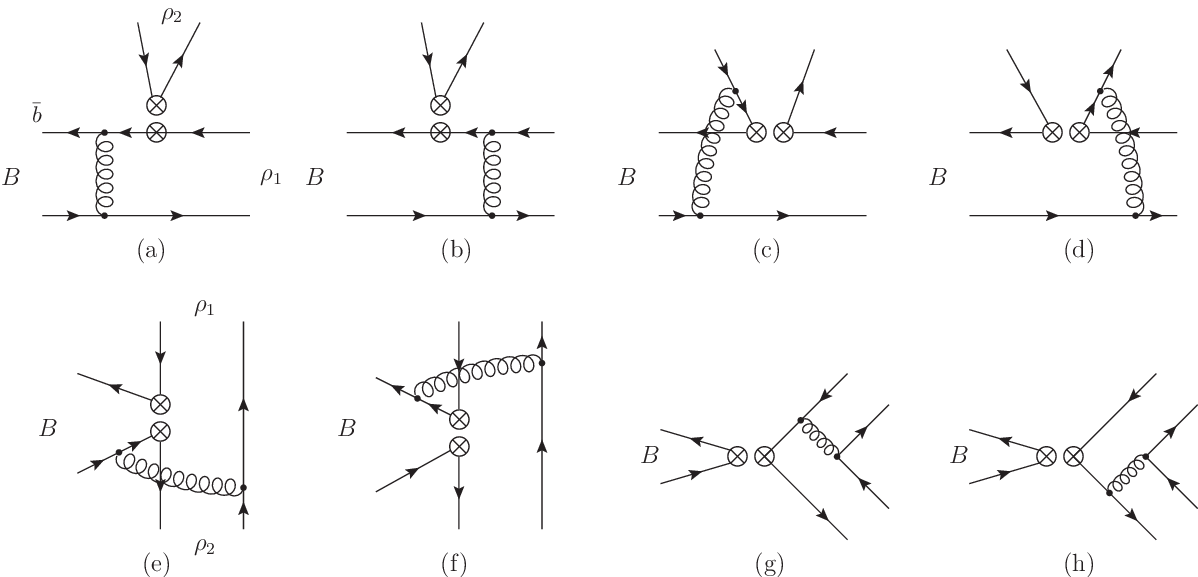}
	\caption{\label{fig:eightdiagram} Decay diagrams contributing to the $B\to\rho\rho$ decays. 
		Diagrams (a, b) are factorizable emission diagrams, (c, d) are nonfactorizable emission diagrams,
		(e, f) are nonfactorizable annihilation diagrams, and (g, h) are factorizable annihilation diagrams.}
\end{figure}

Considering the mass of the $B$ meson is much heavier than the final state $\rho$ mesons, the two outgoing $\rho$ mesons are energetic and oppositely aligned
in the center-of-mass frame of the initial $B$ meson.
Through the four-quark operators, the $b$ quark decays into three quarks, where two of them form one $\rho$ meson, and the remaining quark forms the other $\rho$
meson with the light (spectator) quark in the $B$ meson.
However, the energy of the spectator quark is in the order of the QCD scale $\Lambda_{\textrm{QCD}}$.
In the framework of PQCD, to compose the fast-moving $\rho$ meson, the spectator quark must be kicked out by exchanging a hard gluon between one of the quarks
in the four-quark operators and the spectator quark at the leading order in QCD.
This process yields four emission diagrams, which are classified into factorizable (a, b) and nonfactorizable (c, d),
according to which quark in the four-quark operators interacts with the hard gluon.
They are shown in figure~\ref{fig:eightdiagram} (a)-(d).

There are other possible topological diagrams.
If the light quark within the $B$ meson also corresponds to the quark in the four-quark operators, the $B$ meson undergoes annihilation
and produces the quark-antiquark pair.
To form the $\rho\rho$ final state, the other pair of quark-antiquark is required.
This pair is generated by the gluon originating from the four-quark operators.
The annihilation topological diagrams are also categorized as factorizable (g, h) and nonfactorizable (e, f) diagrams as depicted in figure~\ref{fig:eightdiagram} (e)-(h).

The $B\to\rho\rho$ decay involves multiple energy scales, including the electroweak scale characterized by the $W$ boson mass $m_W$, $B$ meson scale $m_B$, hard scattering scale
$\sqrt{m_B\Lambda_{\textrm{QCD}}}$, and QCD scale $\Lambda_{\textrm{QCD}}$.
We employ the PQCD approach based on the $k_T$ factorization to separate the physics effects in different energy scales.
The momenta of mesons are defined in the light cone coordinates.
Due to the relatively small mass compared with the momentum of the outgoing $\rho$ meson, the light cone momenta of $\rho$ mesons possess a large component on $+$ (or $-$) direction
and leave the component on the other direction neglected, which simplifies the calculation significantly.
As for the constituent quarks within mesons, their momenta are defined by the parton momentum fractions $x$,
with the transverse momentum $k_T$ in the $\Lambda_{\textrm{QCD}}$ scale included.
The propagators of gluon and quark lead to the dominant term in the amplitude for the diagrams being proportional to the $1/x^3$.
At the end-point region (as $x\to 0$), the $\mathcal{O}(1/x^3)$ dominant term cannot be canceled by the meson distribution amplitudes, resulting in the emergence of the divergence contribution.
In this case, the transverse momentum $k_T$ plays an important role in becoming predominant at the end-point region and eliminating the relevant divergence \cite{Keum:2000ph,Keum:2000wi, Lu:2000em}.
The additional momentum scale $k_T$ induces the extra large logarithms.
To collect these large logarithms up to all orders, the resummation technique is utilized to address these extra large logarithms and produces the Sudakov factor \cite{Li:1992nu}. 
So dose the longitudinal momentum fraction of the partons, which results in large logarithms at the end-point region with $x\to 0$ in the higher order radiative corrections and the resummation of the relevant large logarithms give the threshold factor \cite{Li:2001ay}.

Based on the discussion above, the amplitude for $B\to\rho\rho$ decay at hard scale $t$ can be expressed formally as a convolution
\begin{eqnarray} \label{eq:fac}
	\mathcal{M} &= \int d^3k_0d^3k_1d^3k_2 \Phi^B(\vec{k}_0,t) C(t) 
		H(\vec{k}_0,\vec{k}_1,\vec{k}_2,t) \Phi^{\rho}(\vec{k}_1,t) \Phi^{\rho}(\vec{k}_2,t) \nonumber \\
		&\quad \times \exp\left[-s(P,b)-2\int_{1/b}^t \frac{d\bar{\mu}}{\bar{\mu}}\gamma_\Phi(\alpha_s(\bar{\mu}))\right].
\end{eqnarray}
The Wilson coefficient $C(t)$ implements the evolution from the electroweak scale $m_W$ down to the hard scattering scale $t$.
The hard kernel $H(\vec{k}_0,\vec{k}_1,\vec{k}_2,t)$ is calculable by perturbation theory using the diagrams illustrated in figure~\ref{fig:eightdiagram}.
And the function $\Phi^{B(\rho)}(\vec{k}_{0(1,2)},t)$ represent hadron wave functions, incorporating the contribution down to the QCD scale $\Lambda_{\textrm{QCD}}$.
$\alpha_s(\bar{\mu})$ represents the running coupling of the strong interaction at the energy scale $\bar{\mu}$.
Although the wave functions $\Phi^{B(\rho)}$ cannot be rigorously computed here, they are independent of the specific decay channel and may be investigated by nonperturbative methods
or extracted from the experimental data, thereby enhances the predictive power of the PQCD factorization approach.
The Sudakov factor $s(P,b)$ and the anomalous dimension $\gamma_\Phi$ emerge from the process of resummation.
Their expressions are detailed in the \ref{app:loamp}.

\subsection{Meson distribution amplitudes \label{subsec:lcda}}
As we have mentioned previously, the meson wave functions are important inputs in the study of the nonleptonic $B$ meson decays.
The related parameters are also the main source of uncertainty in the theoretical predictions.
Fortunately, they have been extensively studied in various nonperturbative approaches and constrained by data from experiments \cite{ParticleDataGroup:2024cfk}.

The $B$ meson is treated as a heavy-light two body system.
Its distribution amplitude is defined by the non-local matrix element in the heavy quark effective theory as
\begin{equation}
	\langle 0|\bar{q}_{\gamma}(z)[z,0]b_{\delta}(0)|\bar{B}(\vec{k})\rangle = \int d^3k \Phi_{\delta\gamma}^B(\vec{k},t)\exp(-ik\cdot z),
\end{equation}
where $t$ is the scale for the wave function, $[z,0]$ the Wilson line connecting the space-time coordinates of $b$ quark and the light anti-quark $\bar{q}$, and $\delta$ and $\gamma$ the spinor indices.
In this study, instead of the conventional decomposition form \cite{Beneke:2000wa}, we present the matrix element in another form derived from the QCD-inspired relativistic potential model
that includes the entire spinor structure of the momentum projector \cite{Yang:2011ie,Liu:2013maa,Liu:2015lka, Sun:2016avp}. 
The $B$ meson wave function calculated from the QCD inspired relativistic potential model can be viewed as the wave function at hadronic scale $t=\Lambda_{\mathrm{QCD}}$, which is \cite{Sun:2016avp}
\begin{eqnarray} \label{eq:phib}
	\fl &\Phi_{\delta\gamma}^B(\vec{k},\Lambda_{\mathrm{QCD}}) \nonumber \\
	\fl	&=\frac{-if_Bm_B}{4}K(\vec{k}) \biggl\{(E_Q+m_Q)\frac{1+\not{v}}{2}\biggl[\left(\frac{k_+}{\sqrt{2}}+\frac{m_q}{2}\right)\not{n}_+ 
		+\left(\frac{k_-}{\sqrt{2}}+\frac{m_q}{2}\right)\not{n}_{-}-k_\perp^\mu\gamma_\mu\biggr]\gamma_5 \nonumber \\
	\fl	&\quad-(E_q+m_q)\frac{1-\not{v}}{2}\biggl[\left(\frac{k_+}{\sqrt{2}}-\frac{m_q}{2}\right)\not{n}_+ 
		+\left(\frac{k_-}{\sqrt{2}}-\frac{m_q}{2}\right)\not{n}_--k_\perp^\mu\gamma_\mu\biggr]\gamma_5\biggr\}_{\delta\gamma},
\end{eqnarray}
where $f_B$ represents the decay constant of the $B$ meson, $n_\pm$ are light-like vectors defined as $n_\pm ^\mu = (1,0,0,\mp 1)$, $v$ is the $B$ meson four speed,
and the subscripts $Q$ and $q$ denote the heavy and light quarks contained within the $B$ meson respectively.
The momentum $k$ of the component quark is expressed in the light-cone coordinates, decomposed into $k_\pm=(k^0\pm k^3)/\sqrt{2}$ and $k_\perp^\mu=(0,k_1,k_2,0)$.
The function $K(\vec{k})$ is proportional to the momentum space wave function $\Psi(\vec{k})$ as follows \cite{Sun:2016avp}
\begin{eqnarray}
	\fl K(\vec{k}) &= \frac{2N_B\Psi_0(\vec{k})}{\sqrt{E_qE_Q(E_q+m_q)(E_Q+m_Q)}} 
		= \frac{2N_Ba_1 \exp\left(a_2 |\vec{k}|^2 + a_3 |\vec{k}| +a_4\right)}{\sqrt{E_qE_Q(E_q+m_q)(E_Q+m_Q)}},
\end{eqnarray}
with the normalization factor $N_B=\sqrt{{3}/\left[(2\pi)^3m_B\right]}/f_B$.
The parameters $a_i\ (\mathrm{where } i=1,2,3,4)$ with combined uncertainty in the wave function are fitted from the numerical result of the $B$ meson wave function calculated in the relativistic potential model as \cite{Sun:2016avp}
\begin{eqnarray} \label{eq:a1234}
	a_1 &= 4.55_{-0.30}^{+0.40}~\mathrm{GeV}^{-3/2}, & a_2 = -0.39_{-0.20}^{+0.15}~\mathrm{GeV}^{-2}, \nonumber\\
    a_3 &= -1.55\pm 0.20~\mathrm{GeV}^{-1}, & a_4 = -1.10_{-0.05}^{+0.10}.
\end{eqnarray}
Based on this wave function, the renormalization-group evolution of the matrix element have been discussed in ref.~\cite{Lu:2021gwt}. 
The evolution effect related to the $B$ meson wave function can all be attributed to the renormalization effect of quark and antiquark in the meson, which can be collected into the evolution factor associated together with the Sudakov factor for $B$ meson.

For the $\rho$ meson, the distribution amplitudes are a bit more complicated than those of pseudoscalar mesons, such as the pion, due to the inclusion of the different polarization states.
The decay constants of the $\rho$ meson $f_\rho^{\parallel,\perp}$ are defined by \cite{Ball:1998ff}
\begin{eqnarray} \label{eq:rhof}
	\langle 0 |\bar{u}(0) \gamma_\mu d(0) | \rho^-(p,\epsilon)\rangle &= f_\rho^\parallel m_\rho \epsilon_\mu, \nonumber \\
	\langle 0 |\bar{u}(0) \sigma_{\mu\nu} d(0) | \rho^-(p,\epsilon)\rangle &= if_\rho^\perp \left(\epsilon_\mu p_\nu - \epsilon_\nu p_\mu\right),
\end{eqnarray}
where $\epsilon_\mu$ is the meson polarization vector.
Taking into account the Fierz identity and the definitions of the light-cone distribution amplitude (LCDA), 
the non-local matrix element for the vacuum to the $\rho$ meson can be expressed in the twist expansion as \cite{Ball:1998ff, Ball:1998sk, Kurimoto:2001zj, Ball:2007rt} 
\begin{eqnarray} \label{eq:phirhol}
	\fl \langle  \rho^-&(p,\epsilon_L)|\bar{d}_\xi(y)u_\eta(0)|0\rangle 
		=\frac{1}{\sqrt{2N_c}}\int dx d^2k_{q\perp} \exp\left[i(xp\cdot y - k_{q\perp}\cdot y_\perp)\right] \nonumber \\
	\fl	&\quad\times\left[m_{\rho}\not{\epsilon_L^*}\phi_{\rho}(x,k_{q\perp}) + \not{\epsilon_L^*}\not{p}\phi_{\rho}^t(x,k_{q\perp})
		+ m_{\rho}\phi_{\rho}^s(x,k_{q\perp})\right]_{\eta\xi},
\end{eqnarray}
\begin{eqnarray} \label{eq:phirhot}
	\fl \langle  \rho^-&(p,\epsilon_T)|\bar{d}_\xi(y)u_\eta(0)|0\rangle 
		= \frac{1}{\sqrt{2N_c}}\int dx d^2k_{q\perp} \exp\left[i(xp\cdot y - k_{q\perp}\cdot y_\perp)\right] \nonumber \\
	\fl	&\quad \times \biggl[\epsilon_T^* \not{p} \phi_\rho^T(x, k_{q\perp}) + m_\rho \epsilon_T^* \phi_\rho^v(x, k_{q\perp})  
		-i m_\rho \epsilon_{\mu\nu\omega\sigma} \gamma_5 \gamma^\mu \epsilon_T^{*\nu} \frac{p^\omega n^\sigma}{p\cdot n} \phi_\rho^a(x, k_{q\perp})\biggr]_{\eta\xi},
\end{eqnarray}
where $N_c$ is the color factor and $n$ a light-like vector with its direction opposite to that of the momentum of $\rho$ meson, i.e., $n^{\mu}=(1,0,0,-1)$ or $n^{\mu}=(1,0,0,+1)$ for $\rho$ moving in the plus or minus direction of $z$-axis, respectively.
The subscripts of the polarization vector $\epsilon^*$ denote the longitudinal (L) and the transverse (T) polarizations, respectively.
And one should not confuse the polarization vector with the Levi-Civita tensor $\epsilon_{\mu\nu\omega\sigma}$. 
The transverse momentum of the light quark is denoted by $k_{q\perp}$, with an additional subscript $q$ to distinguish it from the case of $B$ meson.
For simplicity, the detailed expressions of the twist-2 $\phi_\rho^{(T)}$ and twist-3 LCDAs $\phi_\rho^{t(s,v,a)}$ are organized in \ref{app:rholcda}.

\subsection{Leading order amplitudes \label{subsec:loamp}}
By combining the factorization expression in eq.~(\ref{eq:fac}), the effective Hamiltonian in eq.~(\ref{eq:heff}), and the meson wave functions in eqs.~(\ref{eq:phib}), (\ref{eq:phirhol}), and (\ref{eq:phirhot}), we are now prepared to calculate the leading order diagrams shown in figure~\ref{fig:eightdiagram}.
As specified previously, the diagrams in figure~\ref{fig:eightdiagram} have been classified into four types: factorizable emission, nonfactorizable emission,
factorizable annihilation, and nonfactorizable annihilation diagrams.
Those resulting expressions are represented by the following conventions: the main symbol indicates whether the diagram is factorizable ($F$) or nonfactorizable ($\mathcal{M}$),
the subscript denotes emission ($e$) or annihilation ($a$), and the superscripts ($\_,R,P$) correspond to the three types of inserted four-quark operators $(V-A)(V-A)$, $(V-A)(V+A)$,
and $(S+P)(S-P)$ which comes from the Fierz transformation.
Because of the possible polarization of the final state, the $B\to\rho\rho$ decay is more intricate than that of the $B\to\pi\pi$. 
The subscript is also employed to specify the final state polarization. 
In the linear polarization scheme, the $\rho\rho$ final state have three possible polarization states: both $\rho$ mesons are longitudinally polarized, or both $\rho$ mesons are transversely polarized, either with the parallel or perpendicular polarization vector. 
We denote these three polarization states as $L$, $N$, and $T$. The corresponding decay amplitudes are denoted as $F_{L,N,T}$ for the factorizable diagrams, and $\mathcal{M}_{L,N,T}$ for the nonfactorizable ones. 
The complete results of these amplitudes corresponding to the eight leading order diagrams are presented in \ref{app:loamp}. 

Requiring the outgoing $\rho$ meson to move in the $+$ (or $-$) direction in the light-cone coordinate system, its polarization vector in the transverse direction has two cases.
In the case of both $\rho$ mesons being transversely and parallelly polarized, the polarization vectors can be
\begin{eqnarray}
	\epsilon^{\mu}_{1,2} = \left(0,0,1,0 \right),\quad \mathrm{or},\quad \epsilon^{\mu}_{1,2} = \left(0,0,0,1 \right),
\end{eqnarray}
where the polarization vectors are given in the light-cone coordinate.

Equivalently, in the helicity polarization scheme, the polarization vectors of the final-state $\rho$ mesons may also be written as
\begin{eqnarray}
	\epsilon^{\mu}_\pm = \frac{1}{\sqrt{2}}\left(0,0,1,\pm i\right),
\end{eqnarray}
for $\rho$ meson moving in the ``$+z$" direction, or
\begin{eqnarray}
	\epsilon^{\mu}_\pm = \frac{1}{\sqrt{2}}\left(0,0,1,\mp i\right),
\end{eqnarray}
for $\rho$ meson moving in the ``$-z$" direction. 
In this case, the amplitudes for those diagrams can be expressed as a combination of the linearly polarized amplitudes
\begin{eqnarray}
	F_{e,\pm} = \frac{1}{\sqrt{2}} \left(F_{e,N} \pm i F_{e,T}\right).
\end{eqnarray}
The similar definition can be extended to the amplitudes associated with all the eight diagrams of figure~\ref{fig:eightdiagram}.

\section{Corrections from NLO diagrams, soft form factors, and color-octet contributions \label{sec:cor}}
To improve the precision of the PQCD calculation, we include several next-to-leading-order and nonperturbative contributions that are currently known, in addition to the leading-order diagrams, 
and the strong running coupling $\alpha_s$ at the two-loop order is adopted \cite{ParticleDataGroup:2024cfk}. 
The Wilson coefficients $C(\mu)$ with the NLO QCD corrections, which are derived from the matching of the effective Hamiltonian with the full theory, have been presented in Ref.~\cite{Buchalla:1995vs}.
There are three kinds of significant NLO corrections need to be included at first: the vertex correction, the quark loop correction, and the chromomagnetic penguin correction. 
The effects of these NLO corrections have been extensively discussed \cite{Beneke:1999br, Beneke:2001ev, Mishima:2003wm, Li:2005kt, Yan:2018fif, Chai:2022ptk}. 
Moreover, the soft form factors, which absorb the nonperturbative contributions below the typical hadronic scale of $1~\mathrm{GeV}$, and the contributions from color-octet quark-antiquark components, are also taken into consideration. 
These nonperturbative contributions have been demonstrated to be capable of explaining the longstanding $\pi\pi$ and $K\pi$ puzzles while keeping most of observables for the $B\to PP$ channel consistent with the experimental data \cite{Lu:2022hbp, Wang:2022ihx, Lu:2024jzn}.

\subsection{NLO corrections}
In this work, we include the vertex correction, quark loop, and chromomagnetic penguin contributions into the calculations. 
The NLO corrections for the nonfactorizable diagrams are neglected, because those impacts on the amplitudes are relatively minor compared to the factorizable ones. 
The standard combinations of the Wilson coefficients, which serve to simplify the expressions, are defined as
\begin{eqnarray} \label{eq:Wilsona}
		a_{1(2)}(\mu)=&C_{2(1)}(\mu)+{C_{1(2)}(\mu)}/{N_c}, \nonumber \\
		a_{i(j)}(\mu)=&C_{i(j)}(\mu)+{C_{i+1(j-1)}(\mu)}/{N_c},
\end{eqnarray}
with $i\in\{3,5,7,9\}$ and $j\in\{4,6,8,10\}$.

The vertex corrections have been proven to play a crucial role in reducing the scale dependence of the Wilson coefficients \cite{Li:2005kt}. 
Since these corrections do not lead to the emergence of the end-point divergences within the framework of the collinear factorization theorem, the QCDF results can be directly applied here \cite{Beneke:2001ev}. 
Due to the structure of the vertex correction for a given four-quark operators $O_i$, their contributions can be effectively absorbed into the modification of the Wilson coefficients as follows
\begin{eqnarray} \label{eq:avc}
	a_{1(2)}(\mu)&\to a_{1(2)}(\mu)+\frac{\alpha_s(\mu)}{4\pi}C_F\frac{C_{1(2)}(\mu)}{N_c}V_{1(2)}(\rho), \nonumber\\
    a_{i(j)}(\mu)&\to a_{i(j)}(\mu)+\frac{\alpha_s(\mu)}{4\pi}C_F\frac{C_{i+1(j-1)}(\mu)}{N_c}V_{i(j)}(\rho),
\end{eqnarray}
where the indices $i$ and $j$ are within the same range as what indicated in Eq.~(\ref{eq:Wilsona}), 
and $C_F$ is a color factor, which is defined as $C_F = (N_c^2-1)/2N_c$. 
For the longitudinal component, the function $V(\rho)$ in the NDR scheme is consistent with those presented in Ref.~\cite{Beneke:2003zv} as
\begin{eqnarray}
	\fl &V_{i,L}(\rho)= 
	\left\{ 
		\begin{array}{l}
			12\ln(m_b/\mu) - 18 + \left({2\sqrt{2N_c}}/{f_\rho^\parallel}\right) \int_0^1 dx\phi_\rho(x)g(x), \hspace{2.3em} \mathrm{for }\ i = 1-4,9,10\\
			-12\ln(m_b/\mu) + 6 -\left({2\sqrt{2N_c}}/{f_\rho^\parallel}\right) \int_0^1 dx\phi_\rho(x)g(1-x),\ \mathrm{for }\ i = 5,7 \\
			-\left({2\sqrt{2N_c}}/{f_\rho^\perp}\right) \int_0^1 dx\phi_\rho^s(x)\left[h(x)-6\right], \hspace{7.1em} \mathrm{for }\ i=6,8 \label{eq:vcvL}
		\end{array}
	\right.
\end{eqnarray}
and for the transverse components, $V(\rho)$ is given by \cite{Cheng:2001aa, Yang:2004pm, Zou:2005gw, Li:2006cva}
\begin{eqnarray}
	\fl &V_{i,\pm}(\rho) =
	\left\{
		\begin{array}{l}
			12\ln(m_b/\mu) - 18 + \left({2\sqrt{2N_c}}/{f_\rho^\parallel}\right) \int_0^1dx\left[\phi_\rho^v(x)\pm\phi_\rho^a(x)\right]g(x), \\
			\hspace{20em} \mathrm{for }\ i = 1-4,9,10 \\
			-12\ln(m_b/\mu) + 6 -\left({2\sqrt{2N_c}}/{f_\rho^\parallel}\right) \int_0^1dx\left[\phi_\rho^v(x)\pm\phi_\rho^a(x)\right]g(1-x), \\
			\hspace{20em} \mathrm{for }\ i = 5,7 \label{eq:vcvpm}
		\end{array} 
	\right.
\end{eqnarray}
with the polarization states denoted by the subscripts $L$ and $\pm$. 
Since the transition matrix element of $\rho$ meson to vacuum with the scalar or pseudoscalar operator inserted is zero, $F_{e,L(\pm)}^P=0$, there is no contributions of $V_{6(8),\pm}$ here. 
The hard kernels $g(x)$ and $h(x)$ in the Eqs.~(\ref{eq:vcvL})-(\ref{eq:vcvpm}) are defined by
\begin{eqnarray}
	\fl &g(x) \nonumber \\ 
	\fl &=3\left(\frac{1-2x}{1-x}\ln x-i\pi\right)+\biggl[2\mathrm{Li}_2(x)-\ln^2x 
		+\frac{2\ln x}{1-x}-(3+2i\pi)\ln x-(x\leftrightarrow 1-x)\biggr], \\
	\fl &h(x)=2\mathrm{Li}_2(x)-\ln^2x-(1+2i\pi)\ln x-(x\leftrightarrow 1-x).
\end{eqnarray}

The corrections arising from the virtual quark loop and chromomagnetic penguin operator are contributions with the insertion of the four-quark and chromomagnetic operators in the decay process.
In $B\to\rho\rho$ decays, analogous to the pseudoscalar cases, the effective Hamiltonian for these corrections can be expressed as \cite{Li:2005kt}
\begin{eqnarray} \label{eq:heffqlmp}
	H_{\mathrm{eff}}^{\mathrm{ql}} = &-\sum_{q=u,c,t}\sum_{q^\prime}\frac{G_F}{\sqrt{2}}V_{q} \frac{\alpha_s(\mu)}{2\pi}C^{(q)}(\mu,l^2) 
		\left(\bar{d}\gamma_\mu(1-\gamma_5)T^ab\right) \left(\bar{q}^\prime\gamma^\mu T^aq^\prime\right), \nonumber \\
	H_{\mathrm{eff}}^{\mathrm{mp}} = &-\frac{G_F}{\sqrt{2}}V_{t} C_{8g}(\mu)O_{8g}(\mu).
\end{eqnarray}
The invariant mass of the gluon connecting to the quark loop is denoted by $l^2$. 
The functions $C^{(q)}(\mu,l^2)$ are defined by
\begin{eqnarray}
	\fl &C^{(q)}(\mu,l^2) = 
	\left\{
		\begin{array}{l}
			\left[G^{(q)}(\mu,l^2)-2/3\right]C_2(\mu), \hspace{10em} \mathrm{for }\ q=u,c \\ [0.5em]
			\left[G^{(s)}(\mu,l^2)-2/3\right]C_3(\mu) \\ 
			\qquad +\sum\limits_{q^{\prime\prime}=u,d,s,c}G^{(q^{\prime\prime})}(\mu,l^2)\left[C_4(\mu)+C_6(\mu)\right], \hspace{2.2em} \mathrm{for }\ q=t,
		\end{array} 
	\right.
\end{eqnarray}
where the small contributions from the electroweak penguin operators $O_{7-10}$ are neglected. 
The function $G^{(q)}(\mu,l^2)$ for the $q$ quark loop is given by
\begin{eqnarray}
	&G^{(q)}=-4\int_0^1dxx(1-x)\ln\frac{m_q^2-x(1-x)l^2-i\epsilon}{\mu^2}.
\end{eqnarray}
Taking into account the similar topological structures, the quark loop and chromomagnetic penguin corrections can also be included into the Wilson coefficients $a_{4,6}$ as
\begin{eqnarray}
	a_{4(6)}(\mu)&\to a_{4(6)}(\mu)+\frac{\alpha_s(\mu)}{9\pi}\sum\limits_{q=u,c,t}\frac{V_{q}}{V_{t}}C^{(q)}(\mu,\langle l^2\rangle), \\
	a_{4(6)}(\mu)&\to a_{4(6)}(\mu)-\frac{\alpha_s(\mu)}{9\pi}\frac{2m_B}{\sqrt{\langle l^2\rangle}}C_{8g}^\mathrm{eff}(\mu),
\end{eqnarray}
where the average invariant mass of the virtual gluon $\langle l^2\rangle$ is selected as $m_b^2/4$, which is a reasonable choice for $B$ decays. 
The effective Wilson coefficient $C_{8g}^\mathrm{eff}$ is defined as $C_{8g}^\mathrm{eff}=C_{8g}+C_5$ \cite{Buchalla:1995vs}.

\subsection{Soft transition and production form factors} \label{subsec:softff}

Due to the Lorentz structure of the $B\to\rho$ transition matrix element, form factors $A_{0,1,2,3}(q^2)$ and $V(q^2)$ can be defined as 
\begin{eqnarray} \label{eq:ff}
	\fl \langle  \rho^+&\left(p_\rho, \epsilon\right) |\bar{u} \gamma_\mu (1-\gamma_5) b |\bar{B}^0(p_B) \rangle 
		= \frac{2}{m_B+m_\rho} \epsilon_{\mu\nu\omega\sigma} \epsilon^{*\nu} p_B^\omega p_\rho^\sigma V(q^2)
		+ i\frac{\epsilon^*\cdot q}{q^2} 2m_\rho q_\mu A_0(q^2) \nonumber \\
	\fl &+ i\bigg[\left(m_B+m_\rho\right)  \epsilon_\mu^*  A_1(q^2)  
		- \frac{\epsilon^*\cdot q}{m_B+m_\rho} \left(p_B+p_\rho\right)_\mu A_2(q^2) 
		- \frac{\epsilon^*\cdot q}{q^2} 2m_\rho q_\mu A_3(q^2)\bigg]
		,
\end{eqnarray}
with $q=p_B-p_\rho$ and the relation
\begin{eqnarray}
	A_3(q^2) = \frac{m_B+m_\rho}{2m_\rho} A_1(q^2) - \frac{m_B - m_\rho}{2m_\rho} A_2(q^2).
\end{eqnarray}
Integrating with the definition of the decay constant in Eq.~(\ref{eq:rhof}), we can express the matrix element of the $\bar{B}^0\to\rho^+\rho^-$ decay
within the naive factorization framework \cite{Bauer:1984zv, Bauer:1986bm} as
\begin{eqnarray}
	\fl T &=\langle \rho^-(p_2, \epsilon_2)|\bar{d}\gamma^\mu(1-\gamma_5)u|0\rangle 
		\langle \rho^+(p_1, \epsilon_1)|\bar{u}\gamma_\mu(1-\gamma_5)b|\bar{B}^0(p_B)\rangle \nonumber \\
	\fl &=\frac{2m_\rho f_\rho^\parallel}{m_B+m_\rho} \epsilon_{\mu\nu\omega\sigma} \epsilon_1^{*\mu} \epsilon_2^{*\nu} p_1^\omega p_2^\sigma V(p_2^2) \nonumber \\
	\fl &\quad + i m_\rho f_\rho^\parallel \bigg[\left(m_B+m_\rho\right) \left(\epsilon_1^*\cdot\epsilon_2^*\right) A_1(p_2^2)
		- \frac{2\left(\epsilon_1^*\cdot p_B\right) \left(\epsilon_2^*\cdot p_B\right)}{m_B+m_\rho} A_2(p_2^2) \bigg],
\end{eqnarray}
where the anti-symmetry of the Levi-Civita tensor and the relation $\epsilon_i\cdot p_i = 0$ ($i=1,2$)
have been used in the derivation.
In principle, the form factors should be taken at the value $p_2^2=m^2_\rho$. 
Here, the mass of $\rho$ meson is relatively light. One can approximately neglect its mass and use the values of the form factors at the point where $q^2=0$.

For the three possible polarizations of the final state, the different form factors contribute to the amplitudes of different polarization states. 
The parallel component is contributed by $A_1$, the perpendicular component by $V$, and the longitudinal component by $A_1$ and $A_2$.
In fact, the matrix element $T$ of naive factorization corresponds to the Feynman diagrams shown in Fig.~\ref{fig:eightdiagram} (a) and (b), which is calculated using PQCD approach and are summarized in \ref{app:loamp}. 
As a result, we can obtain the expressions for these form factors in PQCD approach.

By analyzing the contributions from regions of nonperturbative dynamics with scale below the critical scale $\mu_c = 1~\mathrm{GeV}$, as discussed in Ref.~\cite{Lu:2024jzn}, we find the soft contributions are significant in factorizable diagrams. Employing the $B$ meson wave function derived from the relativistic potential model, it is evident that the soft contributions cannot be effectively suppressed by Sudakov factor. The reason is that the suppression to the soft contributions comes from two folds in the calculation of PQCD in general. One is from the direct suppression of the Sudakov factor, the other is the end-point suppression of the meson wave functions. The end-point behavior of $B$ meson wave function used in previous PQCD calculation is about $\sim x^2$ as $x\to 0$, here $x$ is the momentum fraction of the light quark in $B$ meson, which is a strong suppression to the infrared contribution from the end-point region. Here, in this work, the end-point behavior of $B$ meson wave function obtained by solving the wave equation in the potential model is about $\sim \sqrt{x}$ as $x\to 0$. So the suppression of the $B$ meson wave function to the end-point infrared contribution becomes weak, which results in the suppression to the soft contribution being inadequate. 
Therefore, it is necessary to introduce soft form factors corresponding to the emission and annihilation factorizable diagrams to absorb the contributions below the critical scale.
The contributions above the $\mu_c$, which are perturbative, are evaluated using the PQCD approach.
In this case, the $B\to\rho$ transition form factors can be expressed as
\begin{eqnarray} \label{eq:brhoff}
	F^{B\rho}_z = h^{B\rho}_z + \xi^{B\rho}_z,
\end{eqnarray}
where the total form factor, the hard part, and the soft part are denoted by $F$, $h$, and $\xi$, respectively. 
The subscript $z$ runs over all kinds of form factors, i.e. $z\in\{A_0, A_1, A_2, V\}$.
For the factorizable annihilation diagrams illustrated in Fig.~\ref{fig:eightdiagram} (g) and (h), the amplitudes with the $(V-A)(V\pm A)$ four-quark operators inserted are canceled dramatically owing to the identical final-state mesons. 
The dominated contributions originate from the $F_{a,L(N,T)}^P$. 
It is appropriate to define the production form factor through the scalar and pseudoscalar operators as
\begin{eqnarray} \label{eq:pff}
	\langle \rho\rho | (S+P) | 0\rangle = -\frac{1}{2} m_\rho F^{\rho\rho}.
\end{eqnarray}
The production form factor can also be separated by the critical scale as transition ones
\begin{eqnarray}
	F^{\rho\rho} = h^{\rho\rho} + \xi^{\rho\rho}.
\end{eqnarray}

The critical scale $\mu_c$ may be slightly adjusted around the $1~\mathrm{GeV}$, which does not significantly affect the theoretical results \cite{Wang:2022ihx}.
These factors may also absorb the potential contributions from higher-order diagrams.
With the inclusion of soft factors, the reliability of the PQCD approach is improved.
However, the cost to separate the hard and soft contributions is the loss of the ability to completely calculate the form factor within the PQCD framework,
leading to the form factors being treated as input parameters.
Transition form factors have been investigated thoroughly with several nonperturbative methods in the literature \cite{Kurimoto:2001zj, Ball:2004rg, Khodjamirian:2006st,Ivanov:2007cw, Ivanov:2011aa, Bharucha:2015bzk, Cheng:2018ouz, Gubernari:2018wyi, Gao:2019lta}. 
We can refer to these results in the numerical calculations.
For the production form factors, there is no enough information from both experimental measurements and theoretical studies up to now. 
Therefore, they are treated as unknown input parameters with their values being fitted based on the experimental data of $B$ decays in this work.

\subsection{Color-octet contributions}
The quark-antiquark pairs that form mesons in hadronic production process are usually considered to be in color-singlet states in general, since hadrons must be colorless. 
However, it is possible that the quark-antiquark pairs produced in the hard interaction scale are in color-octet state before they hadronize into mesons in final state, and then the color-octet quark-antiquark pairs transfer into color-singlet states by exchanging soft gluons in long-distance scale. 
In our previous works in Refs.~\cite{Lu:2022hbp,Wang:2022ihx,Lu:2024jzn}, the color-octet contributions are successfully used to solve the $\pi\pi$, $K\pi$ puzzles and analyze all the other $B\to PP$ decays. 
In this work we continue to consider the contribution of color-octet quark-antiquarks in $B\to \rho\rho$ decays. 
To take the color-octet contributions into account, we can consider the case that the quark-antiquark pairs in the final state in Fig.~\ref{fig:eightdiagram} are in color non-singlet state. 
By using the relation of color $\mathrm{SU(3)}_c$ generators 
\begin{eqnarray}
	T^a_{ik}T^a_{jl}=-\delta_{ik}\delta_{jl}/{2N_c}+\delta_{il}\delta_{jk}/2,
\end{eqnarray}
we can separate the contributions with final quark-antiquark pairs in color-singlet and color-octet states. 
The color-singlet contributions are the usual ones that calculated in PQCD approach, while the color-octet contributions are the extra contributions that we need to consider here. 
The details to extract the color-octet contributions can be found in Refs.~\cite{Wang:2022ihx,Lu:2024jzn}. 
The amplitudes for the final quark-antiquark pairs in color-octet state for the eight diagrams shown in Fig.~\ref{fig:eightdiagram} can be related to the color-singlet ones, and the results are given by
\begin{eqnarray} \label{eq:fm8}
	\fl F_{e,L(N,T)}^{(P,R),8} &\equiv 2N_c^2 F_{e,L(N,T)}^{(P,R),a} - \frac{N_c}{C_F} F_{e,L(N,T)}^{(P,R),b}, \ 
    \mathcal{M}_{e,L(N,T)}^{(P,R),8} \equiv 2 N_c^2 \mathcal{M}_{e,L(N,T)}^{(P,R),c} - \frac{N_c}{C_F} \mathcal{M}_{e,L(N,T)}^{(P,R),d}, \nonumber \\
    \fl \mathcal{M}_{e,L(N,T)}^{(P,R),8^\prime} &\equiv \frac{N_c^2}{C_F} \mathcal{M}_{e,L(N,T)}^{(P,R)}, \ 
	F_{a,L(N,T)}^{(P,R),8} \equiv -\frac{N_c^2}{C_F} F_{a,L(N,T)}^{(P,R)}, \ 
	\mathcal{M}_{a,L(N,T)}^{(P,R),8} \equiv -\frac{N_c}{C_F} \mathcal{M}_{a,L(N,T)}^{(P,R)},
\end{eqnarray}
where the superscripts $8$ and $8^\prime$ denote the expressions corresponding to the color-octet contributions, and the symbols $a,b,c$, and $d$ indicate the diagrams from which the amplitudes are obtained. 
In the derivation of the amplitudes for the color-octet contributions, we have assumed that the momentum distribution of the color-octet quark-antiquark pair is the same as that of the corresponding color-singlet pair in a meson. 
However, the transformation from color-octet to color-singlet states is essentially of long-distance dynamics which cannot be reliably calculated within perturbation theory. 
Therefore, the parameters $Y^8_{F,M}$ are introduced to describe this transformation procedure, where $F$ and $M$ represent the factorizable and nonfactorizable amplitudes, respectively. 
For the sake of simplicity and to avoid introducing excessive number of parameters, we do not distinguish different polarization states for the color-octet parameters here. 
The entire process, demonstrated with a specific example, has been comprehensively explained in Refs.~\cite{Wang:2022ihx,Lu:2024jzn}. 

Incorporating all the corrections, the $B\to\rho\rho$ decay amplitudes can be expressed as follows:
\begin{eqnarray}
	\fl \sqrt{2}&\mathcal{M}\left(B^-\to \rho^-\rho^0\right)_\sigma \nonumber \\
		\fl	&=V_u \Big[\left(a_1+a_{1,VC}+a_2+a_{2,VC}\right) F_{e,\sigma}
			- 2 \eta_\sigma \left(a_1^{\mu_c}+a_{1,VC}^{\mu_c}+a_2^{\mu_c}+a_{2,VC}^{\mu_c}\right) f_\rho^\parallel \xi^{B\rho}_\sigma \nonumber \\
		\fl	& \quad+ \left(C_1+C_2\right) F_{e,\sigma}^8 Y_F^8
			+ \left(C_1 + C_2\right)\mathcal{M}_{e,\sigma}/N_c + \left(C_1 + C_2\right)\left(\mathcal{M}_{e,\sigma}^8 + \mathcal{M}_{e,\sigma}^{8^\prime}\right) Y_M^8 \Big] \nonumber \\
		\fl & \quad-V_t\left(3/2\right)\Big\{\left(a_7+a_{7,VC}+a_9+a_{9,VC}+a_{10}+a_{10,VC}\right) F_{e,\sigma}
			- 2 \eta_\sigma \left(a_7^{\mu_c}+a_{7,VC}^{\mu_c} \right. \nonumber \\ \fl & \quad \left.+a_9^{\mu_c}+a_{9,VC}^{\mu_c}+a_{10}^{\mu_c}+a_{10,VC}^{\mu_c}\right) f_\rho^\parallel \xi^{B\rho}_\sigma
			+ \left(C_8 +C_9+C_{10}\right) F_{e,\sigma}^8 Y_F^8 
			+ \left(1/N_c\right) \nonumber \\ \fl & \quad \times \left[\left(C_9+C_{10}\right) \mathcal{M}_{e,\sigma}
			+ C_7 \mathcal{M}_{e,\sigma}^R + C_8 \mathcal{M}_{e,\sigma}^P \right]
			+ \Big[\left(C_9+C_{10}\right) \left(\mathcal{M}_{e,\sigma}^8 + \mathcal{M}_{e,\sigma}^{8^\prime}\right) \nonumber \\
		\fl	& \quad+ C_7 \mathcal{M}_{e,\sigma}^{R,8} + C_8\mathcal{M}_{e,\sigma}^{R,8^\prime}
			+ C_8 \mathcal{M}_{e,\sigma}^{P,8} + C_7\mathcal{M}_{e,\sigma}^{P,8^\prime}\Big] Y_M^8 \Big\}, \label{eq:rho-rho0amp} 
\end{eqnarray}
\begin{eqnarray}
	\fl	&\mathcal{M}\left(\bar{B}^0 \to \rho^+\rho^-\right)_\sigma \nonumber \\
		\fl &=V_u \Big[\left(a_1+a_{1,VC}\right) F_{e,\sigma} - 2 \eta_\sigma \left(a_1^{\mu_c}+a_{1,VC}^{\mu_c}\right) f_\rho^\parallel \xi^{B\rho}_\sigma
			+ a_2 F_{a,\sigma}
			+ \left(C_1 F_{e,\sigma}^8 + a_2 F_{a,\sigma}^8\right) Y_F^8 \nonumber \\
		\fl &\quad+ \left(1/N_c\right) \left(C_1 \mathcal{M}_{e,\sigma} + C_2 \mathcal{M}_{a,\sigma}\right)
			+ \left(C_1 \mathcal{M}_{e,\sigma}^8 + C_2 \mathcal{M}_{e,\sigma}^{8^\prime} + C_2 \mathcal{M}_{a,\sigma}^8\right) Y_M^8 \Big] 
			- V_t \Big\{\left(a_4 \right. \nonumber \\ \fl&\quad\left. +a_{4,VC}+a_{4,QL}+a_{4,MP}+a_{10}+a_{10,VC}\right)F_{e,\sigma}
			- 2 \eta_\sigma \left(a_4^{\mu_c}+a_{4,VC}^{\mu_c}+a_{4,QL}^{\mu_c}+a_{4,MP}^{\mu_c} \right.\nonumber \\ \fl &\quad\left.+a_{10}^{\mu_c}+a_{10,VC}^{\mu_c}\right) f_\rho^\parallel \xi^{B\rho}_\sigma
			+ \left(2a_3+a_4+a_9/2-a_{10}/2\right) F_{a,\sigma} + \left(2a_5+a_7/2\right) F_{a,\sigma}^R \nonumber \\
		\fl &\quad + \left(a_6-a_8/2\right) F_{a,\sigma}^P -2 r_\rho \chi_B f_B \left(a_6^{\mu_c}-a_8^{\mu_c}/2\right) \xi^{\rho\rho} 
			+ \left[\left(C_3+C_9\right) F_{e,\sigma}^8 
			+ \left(2a_3+a_4 \right.\right. \nonumber \\ \fl &\quad \left.\left.+a_9/2-a_{10}/2\right) F_{a,\sigma}^8 + \left(2a_5+a_7/2\right) F_{a,\sigma}^{R,8} + \left(a_6-a_8/2\right) F_{a,\sigma}^{P,8}\right] Y_F^8 
			+ \left(1/N_c\right) \nonumber \\ \fl &\quad\times\left[\left(C_3+C_9\right) \mathcal{M}_{e,\sigma} 
			+ \left(C_5+C_7\right) \mathcal{M}_{e,\sigma}^R
			+ \left(C_3+2C_4-C_9/2+C_{10}/2\right) \mathcal{M}_{a,\sigma} \right. \nonumber \\ 
		\fl	&\quad\left.+ \left(C_5-C_7/2\right) \mathcal{M}_{a,\sigma}^R  
			+ \left(2C_6+C_8/2\right) \mathcal{M}_{a,\sigma}^P \right]
			+ \Big[\left(C_3+C_9\right) \mathcal{M}_{e,\sigma}^8 + \left(C_4+C_{10}\right)\mathcal{M}_{e,\sigma}^{8^\prime} \nonumber \\
		\fl	&\quad + \left(C_5+C_7\right) \mathcal{M}_{e,\sigma}^{R,8} + \left(C_6+C_8\right)\mathcal{M}_{e,\sigma}^{R,8^\prime} 
			+ \left(C_3+2C_4-C_9/2+C_{10}/2\right) \mathcal{M}_{a,\sigma}^8 \nonumber \\
		\fl	&\quad + \left(C_5-C_7/2\right) \mathcal{M}_{a,\sigma}^{R,8}
			+ \left(2C_6+C_8/2\right) \mathcal{M}_{a,\sigma}^{P,8} \Big] Y_M^8 \Big\}, \label{eq:rho+rho-amp}
\end{eqnarray}
\begin{eqnarray}
	\fl	\sqrt{2}&\mathcal{M}\left(\bar{B}^0 \to \rho^0\rho^0\right)_\sigma \nonumber \\
		\fl &=V_u \Big[-\left(a_2+a_{2,VC}\right) F_{e,\sigma} + 2 \eta_\sigma \left(a_2^{\mu_c}+a_{2,VC}^{\mu_c}\right) f_\rho^\parallel \xi^{B\rho}_\sigma
			+ a_2 F_{a,\sigma}
			+ \left(-C_2 F_{e,\sigma}^8 + a_2 F_{a,\sigma}^8 \right) \nonumber\\ \fl&\quad\times Y_F^8 
			+ \left(1/N_c\right) \left(-C_2 \mathcal{M}_{e,\sigma}
			+ C_2 \mathcal{M}_{a,\sigma} \right)
			+ \left(- C_2 \mathcal{M}_{e,\sigma}^8 - C_1 \mathcal{M}_{e,\sigma}^{8^\prime} + C_2 \mathcal{M}_{a,\sigma}^8\right) Y_M^8 \Big] \nonumber\\ 
		\fl &\quad -V_t \Big\{\left[a_4+a_{4,VC}+a_{4,QL}+a_{4,MP}-\left(3/2\right)\left(a_7+a_{7,VC}+a_9+a_{9,VC}\right)
			-\left(a_{10}\right.\right.\nonumber\\ \fl&\quad\left.\left.+a_{10,VC}\right)/2\right] F_{e,\sigma}
			- 2 \eta_\sigma \left[a_4^{\mu_c}+a_{4,VC}^{\mu_c}+a_{4,QL}^{\mu_c}+a_{4,MP}^{\mu_c} 
			-\left(3/2\right)\left(a_7^{\mu_c}+a_{7,VC}^{\mu_c} \right.\right.\nonumber\\ \fl&\quad\left.\left.+a_9^{\mu_c}+a_{9,VC}^{\mu_c}\right)
			-\left(a_{10}^{\mu_c}+a_{10,VC}^{\mu_c}\right)/2\right] f_\rho^\parallel \xi^{B\rho}_\sigma
			+ \left(2a_3+a_4+a_9/2-a_{10}/2\right) F_{a,\sigma} \nonumber \\ 
		\fl	&\quad+ \left(2a_5+a_7/2\right) F_{a,\sigma}^R + \left(a_6-a_8/2\right) F_{a,\sigma}^P 
			-2 r_\rho \chi_B f_B \left(a_6^{\mu_c}-a_8^{\mu_c}/2\right) \xi^{\rho\rho} \nonumber \\
		\fl	&\quad + \Big[\left(C_3-3C_8/2-C_9/2-3C_{10}/2\right) F_{e,\sigma}^8  
			+ \left(2a_3+a_4+a_9/2-a_{10}/2\right) F_{a,\sigma}^8 \nonumber \\ 
		\fl &\quad + \left(2a_5+a_7/2\right) F_{a,\sigma}^{R,8} + \left(a_6-a_8/2\right) F_{a,\sigma}^{P,8} \Big] Y_F^8 \nonumber 
			+ \left(1/N_c\right) \Big[\left(C_3-C_9/2-3C_{10}/2\right) \mathcal{M}_{e,\sigma} \nonumber \\
		\fl	&\quad+ \left(C_5-C_7/2\right) \mathcal{M}_{e,\sigma}^R
			- \left(3C_8/2\right) \mathcal{M}_{e,\sigma}^P
			+ \left(C_3+2C_4-C_9/2+C_{10}/2\right) \mathcal{M}_{a,\sigma} \nonumber \\
		\fl	&\quad+ \left(C_5-C_7/2\right) \mathcal{M}_{a,\sigma}^R
			+ \left(2C_6+C_8/2\right) \mathcal{M}_{a,\sigma}^P \Big]
			+ \Big[\left(C_3-C_9/2-3C_{10}/2\right) \mathcal{M}_{e,\sigma}^8 \nonumber \\ 
		\fl &\quad+ \left(C_4-3C_9/2-C_{10}/2\right) \mathcal{M}_{e,\sigma}^{8^\prime}
			+ \left(C_5-C_7/2\right) \mathcal{M}_{e,\sigma}^{R,8} + \left(C_6-C_8/2\right) \mathcal{M}_{e,\sigma}^{R,8^\prime} \nonumber \\
		\fl	&\quad- \left(3C_8/2\right) \mathcal{M}_{e,\sigma}^{P,8}- \left(3C_7/2\right) \mathcal{M}_{e,\sigma}^{P,8^\prime} 
			+ \left(C_3+2C_4-C_9/2+C_{10}/2\right) \mathcal{M}_{a,\sigma}^8 \nonumber \\ 
		\fl &\quad+ \left(C_5-C_7/2\right) \mathcal{M}_{a,\sigma}^{R,8}
			+ \left(2C_6+C_8/2\right) \mathcal{M}_{a,\sigma}^{P,8} \Big] Y_M^8 \Big\}, \label{eq:rho0rho0amp}
\end{eqnarray}
where the mass ratio $r_\rho$ is defined as the ratio of the masses of the $\rho$ and $B$ mesons, $r_\rho = m_\rho/m_B$, and the normalization factor $\chi_B$ is equal to $1.388$. 
In Eqs.~(\ref{eq:rho-rho0amp})-(\ref{eq:rho0rho0amp}), the subscript $\sigma$, which can be $L$, $N$, and $T$, denotes the polarization configuration of the final state. 
The dimensionless factor $\eta_{L,N,T} = (-1/2r_\rho, r_\rho, -ir_\rho)$ corresponds to the three configurations.
As discussed above in Sec.~\ref{subsec:softff}, the different polarization components are related to the distinct form factors.
The soft transition form factors $\xi^{B\rho}_\sigma$ are defined as $\xi^{B\rho}_{L} = \xi^{B\rho}_{A1}-\xi^{B\rho}_{A2}$,
$\xi^{B\rho}_{N} = \xi^{B\rho}_{A1}$, and $\xi^{B\rho}_{T} = \xi^{B\rho}_V$.
In addition to the standard Wilson coefficients $a$, we employ the $a_{VC}$, $a_{QL}$, and $a_{MP}$ to represent the contributions arising from the vertex correction, quark loop, 
and chromomagnetic penguin diagrams, respectively.
It is essential to emphasize that the multiplication of the Wilson coefficients ($a$ and $C$) with the amplitudes of diagrams ($F$ and $\mathcal{M}$) should be interpreted in the sense of convolution, 
as the relevant energy scale is implied.
On the contrary, the Wilson coefficients $a^{\mu_c}$ at the critical scale $\mu_c$ have been specifically determined and are directly multiplied with the associated soft transition form factors.

\section{Numerical results and discussion \label{sec:num}}
The input parameters for our numerical analysis are summarized below \cite{ParticleDataGroup:2024cfk}:
the $W$ boson mass, $m_W = 80.4~\mathrm{GeV}$;
the masses of the $B$ and $\rho$ mesons, $m_B = 5.279~\mathrm{GeV}$ and $m_\rho = 0.775~\mathrm{GeV}$;
the decay constants of mesons, $f_B = 0.210~\mathrm{GeV}$ \cite{Sun:2019xyw}, $f_\rho^\parallel = 0.216~\mathrm{GeV}$, and $f_\rho^\perp = 0.165~\mathrm{GeV}$ \cite{Ball:2007rt};
the lifetimes of the $B$ mesons, $\tau_{B^\pm} = 1.638~\mathrm{ps}$ and $\tau_{B^0} = 1.517~\mathrm{ps}$; 
and the Fermi constant, $G_F = 1.16638\times 10^{-5}~\mathrm{GeV}^{-2}$. 
Regarding the CKM matrix elements, the Wolfenstein parameterization up to $\mathcal{O}(\lambda^5)$ is employed, and the parameters are fixed as 
$\lambda = 0.22501$, $A = 0.826$, $\bar{\rho} = 0.1591$, and $\bar{\eta} = 0.3523$ \cite{ParticleDataGroup:2024cfk}.

There are also several nonperturbative parameters which encode the contributions below the critical scale $\mu_c$ or describe the transformation from the color-octet to the color-singlet state, which include the soft form factors and the color-octet parameters. 
The $B\to\rho$ transition form factors have been widely studied, and the values obtained from the light-cone sum rule (LCSR) approach are employed in this work \cite{Ball:2004rg, Khodjamirian:2006st, Bharucha:2015bzk, Gubernari:2018wyi}. 
By subtracting the hard part, the soft $B\to\rho$ transition form factors are determined and are presented in Table.~\ref{tab:brhoff}. The uncertainty associated with the hard component arises from the error of the parameters in the meson wave functions, which is about 5\%. The subtraction of the hard part from the total form factor will cause the uncertainty of the soft part of the form factor. Table \ref{tab:brhoff} shows that the uncertainties of the soft form factors are about 20\%$\sim$ 40\%, which is mainly caused by the uncertainties of the total form factors. It is shown in Table \ref{tab:brhoff} that the smaller the soft form factor, the larger the relative uncertainty. 
\begin{table*}
  \caption{\label{tab:brhoff}
    The $B\to\rho$ transition form factors at $q^2=0$.
	The column labeled ``total $F$'' lists the form factors calculated in the LCSR approach.
	The ``hard $h$'' column presents the perturbative contributions above the critical scale $\mu_c$.
	The ``soft $\xi$'' shows the corresponding soft form factors used in the numerical analysis. }
	\begin{indented}
	\item[]\begin{tabular}{@{}cccc}
		\br
		& total $F$ & hard $h$ & soft $\xi$ \\
		\mr
		$A_0$ & $ 0.33\pm 0.05$ & $0.18\pm 0.01$ & $0.14\pm 0.05$ \\
		$A_1$ & $ 0.24\pm 0.03$ & $0.16\pm 0.01$ & $0.08\pm 0.03$ \\
		$A_2$ & $ 0.21\pm 0.04$ & $0.15\pm 0.01$ & $0.06\pm 0.04$ \\
		$V  $ & $ 0.31\pm 0.04$ & $0.15\pm 0.01$ & $0.16\pm 0.04$ \\
		\br
    \end{tabular}
	\end{indented}
\end{table*}

As for the soft production form factor $\xi^{\rho\rho}$ and the color-octet parameters $Y^8_{F,M}$, there is no reliable information in both experiment and theory up to now. 
The constrains imposed by the experimental observables are utilized to extract these parameters. 
The observables of interest include the branching fraction, the $CP$ asymmetry, and the longitudinal polarization fraction, which are defined as
\begin{eqnarray}
	\mathcal{B} &= {\Gamma}/{\Gamma_B} = \frac{G_F^2 m_B^3}{128\pi\Gamma_B}\sum_{\sigma}|\mathcal{M}_\sigma|^2,
\end{eqnarray}
\begin{eqnarray}
	A_{CP} &= \left({\overline{\Gamma}-\Gamma}\right)/\left({\overline{\Gamma}+\Gamma}\right),
\end{eqnarray}
and
\begin{eqnarray}
	f_L &= \frac{1}{2} \left(\frac{|\mathcal{M}_L|^2}{\sum_{\sigma}|\mathcal{M}_\sigma|^2} + \frac{|\overline{\mathcal{M}}_L|^2}{\sum_{\sigma}|\overline{\mathcal{M}}_\sigma|^2} \right),
\end{eqnarray}
where $\sigma$ sums over three final-state polarizations $L,N,T$, and $\Gamma_B$ denotes the total decay width of $B$ meson. 

We use the method of Chi-square analysis to obtain these nonperturbative parameters. The statistic quantity Chi-square is defined as 
\begin{equation}
\chi^2=\sum_{i=1}^N\frac{(y_i-\mu_i)^2}{\sigma_i^2},
\end{equation}
where $y_i$ is the measured value of any physical quantity, $\mu_i$ the relevant theoretical value of the same quantity, $\sigma_i$ the standard deviation of the experimental data, and $N$ the total number of the treated physical quantities. In the numerical fitting process, $y_i$'s include the branching ratios, $CP$ violations and the longitudinal polarization fractions measured in experiment. 

A typical quantity used in Chi-square analysis is the reduced Chi-square statistic $\chi^2_{\mbox{d.o.f}}$, which is defined by normalizing the Chi-square by the degrees of freedom (DoF)
\begin{equation}
\chi^2_{\mbox{d.o.f}}=\frac{\chi^2}{N-k},
\end{equation}
where $N$ is equivalent to the number of data points measured in experiment, and $k$ the number of the fitted parameters. By means of the global numerical fitting, the reduced Chi-square for the best fit is
\begin{equation}
\chi^2_{\mbox{d.o.f}}=\frac{13.3}{9-6}=4.4,
\end{equation}
and the values of the soft parameters determined in the fitting are 
\begin{eqnarray} \label{eq:fitpara}
	m_\rho F^{\rho\rho} &= 0.268^{+0.024}_{-0.024}\exp\left[\left(0.896^{+0.076}_{-0.076}\right)\pi i\right], \nonumber \\
	Y_F^8 &= 0.127^{+0.017}_{-0.016}\exp\left[\left(-0.281^{+0.063}_{-0.064}\right)\pi i\right], \nonumber \\
	Y_M^8 &= 0.133^{+0.020}_{-0.020}\exp\left[\left(-0.194^{+0.065}_{-0.067}\right)\pi i\right].
\end{eqnarray}
The total production form factor $F^{\rho\rho}$ multiplied by $m_\rho$ is used in the fitting procedure, rather than the soft component $\xi^{\rho\rho}$. The uncertainties of the above soft parameters are obtained in the numerical fitting process. The values of the parameters are assumed to obey Gaussian distribution, as well as the experimental data. The error ranges for the parameters in Eq.~(\ref{eq:fitpara}) correspond to the parameter distribution of $1 \sigma$ range. The uncertainties will propagate to the calculated physical observables in the final result by varying these parameters within the error ranges.

With the parameters outlined above, the numerical results are shown in Table.~\ref{tab:bracpfl}. It is worth to note that the properties of $B\to \rho\rho$ decay given in this table are due to the first calculation in the modified PQCD method with several new ingredients included in it. In order to demonstrate the contributions from various components, the results are systemically organized in the table, which accounts for each contribution. In Table.~\ref{tab:bracpfl}, the columns labeled ``$\mathrm{LO}_{\mathrm{NLOWC}}$'', ``NLO'', ``+$\xi^{B\rho}$'', and ``+$\xi^{B\rho},\xi^{\rho\rho}$'' represent the
results of leading-order diagrams with NLO Wilson coefficients, NLO diagrams incorporating NLO Wilson coefficients, NLO as well as soft transition form factors $\xi^{B\rho}$, and finally, NLO alongside two types of soft form factors, respectively. The ``NLO+Soft'' column includes the NLO diagrams, NLO Wilson coefficients, soft form factors, and color-octet contributions, which are our final results. The uncertainties associated with the numerical calculations stem from three sources: soft form factors, $B$ meson distribution amplitudes, and light meson distribution amplitudes.
The significant uncertainties of the $B\to\rho$ transition form factors constitute the primary source of the uncertainty in the branching fraction.
The experimental data from the PDG \cite{ParticleDataGroup:2024cfk} and HFLAV \cite{HFLAV:2022esi} are listed in the last column for comparative analysis.
\begin{table*}
	\caption{\label{tab:bracpfl} Branching fraction $(\times 10^{-6})$, \textit{CP} asymmetry, and longitudinal polarization fraction of the $B\to \rho\rho$ decay.}
    \footnotesize
	\begin{tabular}{@{}ccccccccc}
		\br
		Mode & $\mathrm{LO}_{\mathrm{NLOWC}}$ & NLO & $+\xi^{B\rho}$ & $+\xi^{B\rho},\xi^{\rho\rho}$ & NLO+Soft & Data \cite{ParticleDataGroup:2024cfk, HFLAV:2022esi} \\
		\mr
		$\mathcal{B}(B^+\to \rho^+\rho^0)$	&7.4   &6.7   &12.0  &12.0  &$22.9  ^{+5.8  +0.8   +2.2  }_{-5.4  -1.7   -2.3  }$ &$24.0  \pm 1.9 $  \\
		$\mathcal{B}(B^0\to \rho^+\rho^-)$	&10.7  &12.0  &24.4  &22.0  &$30.6  ^{+10.9 +0.6   +2.0  }_{-9.6  -1.3   -1.8  }$ &$27.7  \pm 1.9 $  \\
		$\mathcal{B}(B^0\to \rho^0\rho^0)$	&0.30  &0.06  &0.06  &0.39  &$0.98  ^{+0.27 +0.06  +0.51 }_{-0.25 -0.11  -0.36 }$ &$0.96  \pm 0.15$  \\  [0.5em]
		$A_{CP}(B^+\to \rho^+\rho^0)$  		&0.00  &0.00  &0.00  &0.00  &$0.01  ^{+0.01 +0.00  +0.01 }_{-0.01 -0.00  -0.00 }$ &$-0.05 \pm 0.05$  \\
		$A_{CP}(B^0\to \rho^+\rho^-)$  		&-0.03 &-0.08 &-0.08 &0.05  &$-0.01 ^{+0.04 +0.00  +0.01 }_{-0.03 -0.01  -0.01 }$ &$ 0.00 \pm 0.09$  \\
		$A_{CP}(B^0\to \rho^0\rho^0)$  		&0.2   &0.8   &0.9   &0.5   &$0.3   ^{+0.2  +0.0   +0.1  }_{-0.2  -0.0   -0.1  }$ &$-0.2  \pm 0.9 $  \\  [0.5em]
		$f_L(B^+\to \rho^+\rho^0)$  		&0.969 &0.970 &0.938 &0.938 &$0.972 ^{+0.009+0.000 +0.003}_{-0.010-0.001 -0.005}$ &$0.950 \pm 0.016$ \\
		$f_L(B^0\to \rho^+\rho^-)$  		&0.921 &0.920 &0.888 &0.879 &$0.940 ^{+0.025+0.000 +0.005}_{-0.031-0.002 -0.007}$ &$0.990 \pm 0.020$ \\
		$f_L(B^0\to \rho^0\rho^0)$  		&0.84  &0.83  &0.77  &0.47  &$0.76  ^{+0.04 +0.02  +0.07 }_{-0.05 -0.02  -0.11 }$ &$0.71  \pm 0.06$  \\
		\br
	\end{tabular}
\end{table*}

In the ``$\mathrm{LO}_{\mathrm{NLOWC}}$'' column, the branching fractions are considerably lower than the experimental results, amounting to only about one-third of the observed values. 
The inclusion of corrections from the NLO diagrams does not significantly affect the theoretical predictions, except for the $\rho^0\rho^0$ channel, which is suppressed at leading order due to the cancellation of Wilson coefficients in the tree level and exhibits sensitivity to the NLO corrections. 
The comparison of the columns ``NLO" and ``$\mathrm{LO}_{\mathrm{NLOWC}}$'' indicates that the reliability of the perturbative calculation is improved by the introduction of the soft infrared cutoff at the critical scale $\mu_c$. 
The soft transition form factor $\xi^{B\rho}$, associated with the emission diagrams, almost doubles the branching fractions in the $\rho^+\rho^0$ and $\rho^+\rho^-$ channels. 
Additionally, it reduces the longitudinal polarization fractions, suggesting that the transverse polarization amplitudes are enhanced more than the longitudinal ones. 
However, the soft production form factor, which absorbs the nonperturbative contributions related to the annihilation diagrams, substantially influences the $\rho^0\rho^0$ channel, while impacting the other channels slightly. It is interesting to know how sensitive the physical results are to the fitted production form factors. We checked this by varying the fitted production form factor about 5\%, and we find that the corresponding branching ratios change by less than $4$\%, which indicates that the physical results are not so sensitive to the fitted values. 

\begin{table}
  \caption{\label{tab:varymuc}
	The variations of the numerical results by employing the slightly different infrared cutoff scale $\mu_c$, as the total transition form factor, the production form factor, and the contribution from the color-octet operator are kept invariant.}
	\footnotesize
	\begin{tabular}{@{}cccccccccccc}
		\br
		Mode & $0.9~\mathrm{GeV}$ & $1.0 ~\mathrm{GeV}$  & $1.1~\mathrm{GeV}$   & $1.3~\mathrm{GeV}$   & $1.5~\mathrm{GeV}$   & $2.0~\mathrm{GeV}$   & Data \cite{ParticleDataGroup:2024cfk, HFLAV:2022esi} \\
		\mr
		$\mathcal{B}(B^+\to \rho^+\rho^0)$	& $23.3 $ & $22.9 $ & $22.6 $ & $22.4 $ & $21.7 $ & $20.8 $ & $24.0  \pm 1.9 $  \\
		$\mathcal{B}(B^0\to \rho^+\rho^-)$	& $32.1 $ & $30.6 $ & $29.5 $ & $28.4 $ & $27.5 $ & $27.2 $ & $27.7  \pm 1.9 $  \\
		$\mathcal{B}(B^0\to \rho^0\rho^0)$	& $1.10 $ & $0.98 $ & $0.93 $ & $0.89 $ & $0.85 $ & $0.73$ & $0.96  \pm 0.15$  \\ [0.5em]
		$A_{CP}(B^+\to \rho^+\rho^0)$  		& $0.01 $ & $0.01 $ & $0.01 $ & $0.01 $ & $0.01 $ & $0.01 $ & $-0.05 \pm 0.05$  \\
		$A_{CP}(B^0\to \rho^+\rho^-)$  		& $-0.02$ & $-0.01$ & $-0.01$ & $0.00 $ & $0.00 $ & $0.00 $ & $ 0.00 \pm 0.09$  \\
		$A_{CP}(B^0\to \rho^0\rho^0)$  		& $0.3  $ & $0.3  $ & $0.3  $ & $0.3  $ & $0.3  $ & $0.3  $ & $-0.2  \pm 0.9 $  \\ [0.5em]
		$f_L(B^+\to \rho^+\rho^0)$  		& $0.972$ & $0.972$ & $0.971$ & $0.970$ & $0.969$ & $0.965$ & $0.950 \pm 0.016$ \\
		$f_L(B^0\to \rho^+\rho^-)$  		& $0.936$ & $0.940$ & $0.942$ & $0.945$ & $0.945$ & $0.948$ & $0.990 \pm 0.020$ \\
		$f_L(B^0\to \rho^0\rho^0)$  		& $0.69 $ & $0.76 $ & $0.81 $ & $0.87 $ & $0.88 $ & $0.84 $ & $0.71  \pm 0.06$  \\
		\br
    \end{tabular}
\end{table}
The other interesting point worth to check is how the branching ratios and $CP$ violations change with the value of the critical infrared cutoff scale $\mu_c$. Table.~\ref{tab:varymuc} presents the numerical results by varying the value of $\mu_c$ from 0.9 to 2.0 GeV. As shown in Table \ref{tab:varymuc}, the theoretical results are only slightly changed as $\mu_c$ varying within the range from $0.9~\mathrm{GeV}$ to $1.3~\mathrm{GeV}$, with the deviations of the branching ratios being less than a few percent. The variations of $A_{CP}$ and $f_L$ with this range are also very small. As $\mu_c$ is enlarged up to 2 GeV, the deviation of the physical result begins to be apparent. Therefore, phenomenologically selecting $\mu_c=1~\mathrm{GeV}$ as the separation scale of the perturbative and nonperturbative contributions is a reasonable choice. Certainly, the value of $\mu_c$ can be slightly changed around 1 GeV in practice. 

At last, the color-octet contributions, which play an essential part, improve the branching fractions and longitudinal polarization fractions, aligning them more closely with the experimental data. 
Within the bounds of uncertainty, the theoretical calculations correspond well with the experimental values. 

In light of the fact that nonperturbative parameters in Eq.~(\ref{eq:fitpara}) are fitted from the experimental data, one might doubt whether the approach is capable of addressing the decay channels for which our understanding is still limited. Fortunately, the various decay channels can be interrelated through the SU(3) symmetry. The corresponding parameters may be correlated by the SU(3) symmetry and symmetry-breaking parameters, which has been achieved in the $B\to PP$ decays \cite{Lu:2024jzn}. The approach used in this work can not only be successful in explaining the experimental data for all $B\to PP$ decay modes \cite{Lu:2024jzn}, but also can be extended to $B\to PV$ and $VV$ decays, which have been underway. Moreover, the parameters employed in this work closely resemble those in $B\to PP$ decays. There exists a promising opportunity to unify all charmless two-body decays of the $B$ meson within a single framework, thereby improving the predictive power of the approach.

\section{Summary \label{sec:sum}}
In this work, the $B\to\rho\rho$ decay is examined within the framework of the modified PQCD approach. 
Employing the distribution amplitudes of $B$ meson derived from the relativistic potential model, the implementation of the infrared cutoff at the critical scale $\mu_c$ is necessary to enhance the reliability of the perturbative calculations. 
The contributions below the critical scale are characterized by soft form factors, applicable to both emission and annihilation topological diagrams. 
Furthermore, the color-octet contribution and the relevant parameters are found to be crucial for explaining the experimental data of $B\to\rho\rho$ and $\pi\pi$ within a uniform framework. 
Even though the methods for predicting the nonperturbative color-octet parameters in theory require further exploration, this approach possesses the potential to understand the structure of the mechanism for charmless two-body decays of $B$ meson.

\ack{This work is supported in part by the National Natural Science Foundation of China under Contracts No. 12275139, 11875168.}

\section*{Data availability statement}
All theoretical results are included in the paper and summarized in tables, with no additional data. 

\appendix
\section{The leading order amplitudes in PQCD calculation \label{app:loamp}}
With the notations defined in Sec. \ref{subsec:loamp}, we list the results of leading-order diagrams for $B\to\rho\rho$ below. 
For the longitudinal amplitude, the expressions are given by 
\begin{eqnarray} \label{eq:feL}
	\fl F_{e,L}=&-4\sqrt{2N_c}\pi^2\frac{C_F}{N_c}f_Bf_{\rho}^{\parallel}m_B^2 \int_0^{\frac{m_B}{2}} k_{0\perp}dk_{0\perp}\int_{x_0^d}^{x_0^u}dx_0\int_0^1dx_1
		\int_0^{\infty} b_0db_0b_1db_1 K(\vec{k}_0) \nonumber \\
	\fl	&\times\left(\frac{m_B}{2}+\frac{|k_{0\perp}|^2}{2x_0^2m_B}\right) (E_Q+m_Q) 
		J_0(k_{0\perp}b_0)\{\alpha_s(t_e^1)[((x_1-2)E_q-x_1k_0^3)\phi_{\rho}(x_1,b_1) \nonumber \\ \fl &-r_{\rho}((1-2x_1)E_q+k_0^3)(\phi_{\rho}^s(x_1,b_1)-\phi_{\rho}^t(x_1,b_1))] 
		h_e(x_0,1-x_1,b_0,b_1)S_t(x_1) \nonumber \\
	\fl	&\times\exp[-S_B(t_e^1)-S_{\rho}(t_e^1)]+\alpha_s(t_e^2)2r_{\rho}(E_q-k_0^3)\phi_{\rho}^s(x_1,b_1)h_e(1-x_1,x_0,b_1,b_0) \nonumber \\
	\fl	&\times S_t(x_0)\exp[-S_B(t_e^2)-S_{\rho}(t_e^2)]\},
\end{eqnarray}
\begin{eqnarray} \label{eq:meL}
	\fl \mathcal{M}_{e,L}=&-16\pi^2C_Ff_Bm_B^2 \int k_{0\perp}dk_{0\perp}\int dx_0dx_1dx_2 \int b_0db_0b_2db_2
		\left(\frac{m_B}{2}+\frac{|k_{0\perp}|^2}{2x_0^2m_B}\right)  \nonumber\\ \fl&\times K(\vec{k}_0)(E_Q+m_Q)J_0(k_{0\perp}b_0) 
		\phi_{\rho}(x_2,b_2) \{\alpha_s(t_{ne}^1) [-x_2(E_q-k_0^3)\phi_{\rho}(x_1,b_0) \nonumber\\ \fl& -r_{\rho}(1-x_1)(E_q+k_0^3)(\phi_{\rho}^s(x_1,b_0)+\phi_{\rho}^t(x_1,b_0))] 
		h_{ne}(x_0,x_2,1-x_1,b_0,b_2) \nonumber \\
	\fl &\times\exp[-S_B(t_{ne}^1)-S_{\rho}(t_{ne}^1)|_{b_1\rightarrow b_0}-S_{\rho}(t_{ne}^1)]+\alpha_s(t_{ne}^2)[((2-x_1-x_2)E_q \nonumber \\
	\fl	&-(x_1-x_2)k_0^3)\phi_{\rho}(x_1,b_0)+r_{\rho}(1-x_1)(E_q-k_0^3)(\phi_{\rho}^s(x_1,b_0)-\phi_{\rho}^t(x_1,b_0))] \nonumber \\ 
	\fl &\times h_{ne}(x_0,1-x_2,1-x_1,b_0,b_2) \exp[-S_B(t_{ne}^2)-S_{\rho}(t_{ne}^2)|_{b_1\rightarrow b_0}-S_{\rho}(t_{ne}^2)]\},
\end{eqnarray}
\begin{eqnarray} \label{eq:merL}
	\fl \mathcal{M}_{e,L}^R=&-16\pi^2C_Ff_Bm_B^2\int k_{0\perp}dk_{0\perp}\int dx_0dx_1dx_2\int b_0db_0b_2db_2
		\left(\frac{m_B}{2}+\frac{|k_{0\perp}|^2}{2x_0^2m_B}\right) \nonumber \\ \fl &\times K(\vec{k}_0)(E_Q+m_Q)J_0(k_{0\perp}b_0) 
		r_{\rho}\{\alpha_s(t_{ne}^1)[x_2(E_q-k_0^3)\phi_{\rho}(x_1,b_0)(-\phi_{\rho}^s(x_2,b_2) \nonumber\\ \fl&+\phi_{\rho}^t(x_2,b_2)) -r_{\rho}((1-x_1+x_2)E_q+(1-x_1-x_2)k_0^3) 
		(-\phi_{\rho}^s(x_1,b_0)\phi_{\rho}^s(x_2,b_2) \nonumber\\ \fl& +\phi_{\rho}^t(x_1,b_0)\phi_{\rho}^t(x_2,b_2)) -r_{\rho}((1-x_1-x_2)E_q+(1-x_1+x_2)k_0^3)\nonumber \\
	\fl &\times (\phi_{\rho}^t(x_1,b_0)\phi_{\rho}^s(x_2,b_2)-\phi_{\rho}^s(x_1,b_0)\phi_{\rho}^t(x_2,b_2))] h_{ne}(x_0,x_2,1-x_1,b_0,b_2) \nonumber\\ 
	\fl& \times\exp[-S_B(t_{ne}^1)-S_{\rho}(t_{ne}^1)|_{b_1\rightarrow b_0}-S_{\rho}(t_{ne}^1)] 
		+\alpha_s(t_{ne}^2)[(1-x_2)(E_q-k_0^3) \nonumber\\ \fl&\times\phi_{\rho}(x_1,b_0)(\phi_{\rho}^s(x_2,b_2)+\phi_{\rho}^t(x_2,b_2)) -r_{\rho}((2-x_1-x_2)E_q-(x_1-x_2)k_0^3) \nonumber \\
	\fl	&\times (\phi_{\rho}^s(x_1,b_0)\phi_{\rho}^s(x_2,b_2)+\phi_{\rho}^t(x_1,b_0)\phi_{\rho}^t(x_2,b_2)) -r_{\rho}((x_1-x_2)E_q \nonumber\\ \fl&-(2-x_1-x_2)k_0^3) (\phi_{\rho}^t(x_1,b_0)\phi_{\rho}^s(x_2,b_2) +\phi_{\rho}^s(x_1,b_0)\phi_{\rho}^t(x_2,b_2))] \nonumber \\ 
	\fl &\times h_{ne}(x_0,1-x_2,1-x_1,b_0,b_2)\exp[-S_B(t_{ne}^2)-S_{\rho}(t_{ne}^2)|_{b_1\rightarrow b_0}-S_{\rho}(t_{ne}^2)]\},
\end{eqnarray}
\begin{eqnarray} \label{eq:mepL}
	\fl \mathcal{M}_{e,L}^P=&16\pi^2C_Ff_Bm_B^2\int k_{0\perp}dk_{0\perp}\int dx_0dx_1dx_2\int b_0db_0b_2db_2
		\left(\frac{m_B}{2}+\frac{|k_{0\perp}|^2}{2x_0^2m_B}\right)K(\vec{k}_0) \nonumber\\ \fl&\times (E_Q+m_Q) J_0(k_{0\perp}b_0) 
		\phi_{\rho}(x_2,b_2)\{\alpha_s(t_{ne}^1)[((1-x_1+x_2)E_q+(1-x_1-x_2)k_0^3) \nonumber\\ \fl&\times \phi_{\rho}(x_1,b_0) +r_{\rho}(1-x_1)(E_q-k_0^3) 
		(\phi_{\rho}^s(x_1,b_0)-\phi_{\rho}^t(x_1,b_0))] \nonumber\\ \fl&\times h_{ne}(x_0,x_2,1-x_1,b_0,b_2) \exp[-S_B(t_{ne}^1)-S_{\rho}(t_{ne}^1)|_{b_1\rightarrow b_0}-S_{\rho}(t_{ne}^1)] \nonumber \\
	\fl	&+\alpha_s(t_{ne}^2)[-(1-x_2)(E_q-k_0^3)\phi_{\rho}(x_1,b_0) -r_{\rho}(1-x_1)(E_q+k_0^3) \nonumber\\ 
	\fl &\times (\phi_{\rho}^s(x_1,b_0)+\phi_{\rho}^t(x_1,b_0))]\exp[-S_B(t_{ne}^2)-S_{\rho}(t_{ne}^2)|_{b_1\rightarrow b_0}-S_{\rho}(t_{ne}^2)]  \nonumber \\ 
	\fl &\times h_{ne}(x_0,1-x_2,1-x_1,b_0,b_2)\},
\end{eqnarray}
\begin{eqnarray} \label{eq:faL}
	\fl F_{a,L}=&-8\pi C_Ff_Bm_B^2 \int dx_1dx_2 \int b_1db_1b_2db_22\{\alpha_s(t_a^1)[x_2\phi_{\rho}(x_1,b_1)\phi_{\rho}(x_2,b_2) \nonumber \\
	\fl	&-2r_{\rho}^2(1+x_2) \phi_{\rho}^s(x_1,b_1)\phi_{\rho}^s(x_2,b_2) 
		+2r_{\rho}^2(1-x_2)\phi_{\rho}^s(x_1,b_1)\phi_{\rho}^t(x_2,b_2)]S_t(x_2) \nonumber\\ 
	\fl &\times h_a(1-x_1,x_2,b_1,b_2) \exp[-S_{\rho}(t_a^1)-S_{\rho}(t_a^1)]
		+\alpha_s(t_a^2)[-(1-x_1)\phi_{\rho}(x_1,b_1) \nonumber\\ \fl&\times \phi_{\rho}(x_2,b_2) + 2r_{\rho}^2(2-x_1) \phi_{\rho}^s(x_1,b_1)\phi_{\rho}^s(x_2,b_2)
		+2r_{\rho}^2x_1\phi_{\rho}^t(x_1,b_1)\phi_{\rho}^s(x_2,b_2)] \nonumber\\ 
	\fl &\times h_a(x_2,1-x_1,b_2,b_1)S_t(x_1) \exp[-S_{\rho}(t_a^2)-S_{\rho}(t_a^2)]\},
\end{eqnarray}
\begin{eqnarray} \label{eq:fapL}
	\fl F_{a,L}^P=&8\pi C_Ff_Bm_B^2\chi_B\int dx_1dx_2\int b_1db_1b_2db_2 4r_{\rho}
		\{\alpha_s(t_a^1)[-x_2\phi_{\rho}(x_1,b_1)(\phi_{\rho}^s(x_2,b_2) \nonumber\\ \fl& -\phi_{\rho}^t(x_2,b_2)) +2\phi_{\rho}^s(x_1,b_1) 
		\phi_{\rho}(x_2,b_2)] h_a(1-x_1,x_2,b_1,b_2)\exp[-S_{\rho}(t_a^1)-S_{\rho}(t_a^1)] \nonumber\\ \fl&\times S_t(x_2)
		+\alpha_s(t_a^2) [-2\phi_{\rho}(x_1,b_1)\phi_{\rho}^s(x_2,b_2) 
		+ (1-x_1)(\phi_{\rho}^s(x_1,b_1)+\phi_{\rho}^t(x_1,b_1)) \nonumber\\ \fl&\times \phi_{\rho}(x_2,b_2)] 
		h_a(x_2,1-x_1,b_2,b_1)S_t(x_1)\exp[-S_{\rho}(t_a^2)-S_{\rho}(t_a^2)]\},
\end{eqnarray}
\begin{eqnarray} \label{eq:maL}
	\fl	\mathcal{M}_{a,L}=&-16\pi^2C_Ff_Bm_B^2\int k_{0\perp}dk_{0\perp}\int dx_0dx_1dx_2\int b_0db_0b_1db_1
		\left(\frac{m_B}{2}+\frac{|k_{0\perp}|^2}{2x_0^2m_B}\right) \nonumber\\ \fl&\times K(\vec{k}_0)(E_Q+m_Q) J_0(k_{0\perp}b_0) 
		\{\alpha_s(t_{na}^1)[(1-x_1)(E_q+k_0^3)\phi_{\rho}(x_1,b_1)\phi_{\rho}(x_2,b_1) \nonumber \\
	\fl &-r_{\rho}^2((3-x_1+x_2)E_q +(1-x_1-x_2)k_0^3)\phi_{\rho}^s(x_1,b_1) \phi_{\rho}^s(x_2,b_1) \nonumber \\ 
	\fl &-r_{\rho}^2((1-x_1-x_2)E_q-(1+x_1-x_2)k_0^3) \phi_{\rho}^s(x_1,b_1)\phi_{\rho}^t(x_2,b_1) \nonumber\\
	\fl	&+r_{\rho}^2((1-x_1-x_2)+(3-x_1+x_2)k_0^3)\phi_{\rho}^t(x_1,b_1)\phi_{\rho}^s(x_2,b_1) \nonumber\\
	\fl &-r_{\rho}^2((1+x_1-x_2)E_q-(1-x_1-x_2)k_0^3)\phi_{\rho}^t(x_1,b_1)\phi_{\rho}^t(x_2,b_1)] \nonumber \\
	\fl	&\times h_{na}^1(1-x_1,x_2,b_0,b_1)\exp[-S_B(t_{na}^1)-S_{\rho}(t_{na}^1)-S_{\rho}(t_{na}^1)|_{b_2\rightarrow b_1}] \nonumber \\
	\fl &+\alpha_s(t_{na}^2)[-x_2(E_q-k_0^3)\phi_{\rho}(x_1,b_1)\phi_{\rho}(x_2,b_1)
		+r_{\rho}^2((1-x_1+x_2)E_q \nonumber\\ \fl &+(1-x_1-x_2)k_0^3)(\phi_{\rho}^s(x_1,b_1)\phi_{\rho}^s(x_2,b_1)-\phi_{\rho}^t(x_1,b_1)\phi_{\rho}^t(x_2,b_1)) \nonumber \\
	\fl &-r_{\rho}^2((1-x_1-x_2)E_q+(1-x_1+x_2)k_0^3)(\phi_{\rho}^s(x_1,b_1)\phi_{\rho}^t(x_2,b_1) \nonumber\\ \fl& -\phi_{\rho}^t(x_1,b_1)\phi_{\rho}^s(x_2,b_1))] 
		\exp[-S_B(t_{na}^2)-S_{\rho}(t_{na}^2)-S_{\rho}(t_{na}^2)|_{b_2\rightarrow b_1}] \nonumber\\ 
	\fl &\times h_{na}^2(1-x_1,x_2,b_0,b_1)\},
\end{eqnarray}
\begin{eqnarray} \label{eq:marL}
	\fl	\mathcal{M}_{a,L}^R=&16\pi^2C_Ff_Bm_B^2\int k_{0\perp}dk_{0\perp}\int dx_0dx_1dx_2\int b_0db_0b_1db_1
		\left(\frac{m_B}{2}+\frac{|k_{0\perp}|^2}{2x_0^2m_B}\right)K(\vec{k}_0) \nonumber\\ \fl&\times (E_Q+m_Q) J_0(k_{0\perp}b_0) 
		r_{\rho}\{\alpha_s(t_{na}^1)[((2-x_2)E_q+x_2k_0^3)\phi_{\rho}(x_1,b_1)(\phi_{\rho}^s(x_2,b_1) \nonumber\\ \fl&+\phi_{\rho}^t(x_2,b_1)) + ((1+x_1)E_q-(1-x_1)k_0^3) 
		(\phi_{\rho}^s(x_1,b_1)-\phi_{\rho}^t(x_1,b_1))\phi_{\rho}(x_2,b_1)] \nonumber\\ 
	\fl &\times h_{na}^1(1-x_1,x_2,b_0,b_1)\exp[-S_B(t_{na}^1)-S_{\rho}(t_{na}^1)-S_{\rho}(t_{na}^1)|_{b_2\rightarrow b_1}] \nonumber \\
	\fl	&+\alpha_s(t_{na}^2)[x_2(E_q+k_0^3)\phi_{\rho}(x_1,b_1)(\phi_{\rho}^s(x_2,b_1) + \phi_{\rho}^t(x_2,b_1))
		+(1-x_1)(E_q-k_0^3) \nonumber\\ \fl&\times (\phi_{\rho}^s(x_1,b_1)-\phi_{\rho}^t(x_1,b_1))\phi_{\rho}(x_2,b_1)]
		\exp[-S_B(t_{na}^2)-S_{\rho}(t_{na}^2)-S_{\rho}(t_{na}^2)|_{b_2\rightarrow b_1}] \nonumber\\ 
	\fl &\times h_{na}^2(1-x_2,x_3,b_1,b_2)\},
\end{eqnarray}
\begin{eqnarray} \label{eq:mapL}
	\fl \mathcal{M}_{a,L}^P=&-16\pi^2C_Ff_Bm_B^2\int k_{0\perp}dk_{0\perp}\int dx_0dx_1dx_2\int b_0db_0b_1db_1
		\left(\frac{m_B}{2}+\frac{|k_{0\perp}|^2}{2x_0^2m_B}\right) \nonumber\\ \fl&\times K(\vec{k}_0)(E_Q+m_Q) J_0(k_{0\perp}b_0) 
		\{\alpha_s(t_{na}^1)[x_2(E_q-k_0^3)\phi_{\rho}(x_1,b_1)\phi_{\rho}(x_2,b_1) \nonumber\\
	\fl &-r_{\rho}^2((3-x_1+x_2)E_q +(1-x_1-x_2)k_0^3)\phi_{\rho}^s(x_1,b_1)\phi_{\rho}^s(x_2,b_1) \nonumber \\
	\fl &+r_{\rho}^2((1-x_1-x_2)E_q+(3-x_1+x_2)k_0^3) \phi_{\rho}^s(x_1,b_1)\phi_{\rho}^t(x_2,b_1) \nonumber \\
	\fl &-r_{\rho}^2((1-x_1-x_2)E_q -(1+x_1-x_2)k_0^3)\phi_{\rho}^t(x_1,b_1)\phi_{\rho}^s(x_2,b_1) \nonumber \\
	\fl	&-r_{\rho}^2((1+x_1-x_2)E_q-(1-x_1-x_2)k_0^3)\phi_{\rho}^t(x_1,b_1)\phi_{\rho}^t(x_2,b_1)] \nonumber \\ 
	\fl &\times h_{na}^1(1-x_1,x_2,b_0,b_1)
		\exp[-S_B(t_{na}^1)-S_{\rho}(t_{na}^1)-S_{\rho}(t_{na}^1)|_{b_2\rightarrow b_1}] \nonumber \\
	\fl	&+\alpha_s(t_{na}^2)[-(1-x_1)(E_q+k_0^3)\phi_{\rho}(x_1,b_1)\phi_{\rho}(x_2,b_1)
		+r_{\rho}^2((1-x_1+x_2)E_q \nonumber\\ 
	\fl &+(1-x_1-x_2)k_0^3)(\phi_{\rho}^s(x_1,b_1)\phi_{\rho}^s(x_2,b_1)-\phi_{\rho}^t(x_1,b_1)\phi_{\rho}^t(x_2,b_1)) \nonumber \\
	\fl	&+r_{\rho}^2((1-x_1-x_2)E_q+(1-x_1+x_2)k_0^3)(\phi_{\rho}^s(x_1,b_1)\phi_{\rho}^t(x_2,b_1) \nonumber\\ \fl& -\phi_{\rho}^t(x_1,b_1)\phi_{\rho}^s(x_2,b_1))] 
		\exp[-S_B(t_{na}^2)-S_{\rho}(t_{na}^2)-S_{\rho}(t_{na}^2)|_{b_2\rightarrow b_1}] \nonumber\\ 
	\fl &\times h_{na}^2(1-x_1,x_2,b_0,b_1)\}.
\end{eqnarray}

For the parallel amplitude, we have obtained the following results:
\begin{eqnarray} \label{eq:feN}
	\fl F_{e,N}=&-8\sqrt{N_c}\pi^2\frac{C_F}{N_c}f_Bf_{\rho}^{\parallel}m_B^2 \int k_{0\perp}dk_{0\perp}\int dx_0dx_1
		\int b_0db_0b_1db_1 \left(\frac{m_B}{2}+\frac{|k_{0\perp}|^2}{2x_0^2m_B}\right) \nonumber\\ \fl&\times K(\vec{k}_0)(E_Q+m_Q) J_0(k_{0\perp}b_0) 
		r_{\rho2}\{\alpha_s(t_e^1)[(E_q+k_0^3)\phi_{\rho}^T(x_1,b_1)+r_{\rho1}((3-x_1)E_q \nonumber\\ \fl& -(1-x_1)k_0^3)\phi_{\rho}^v(x_1,b_1)
		+r_{\rho1}((1-x_1)E_q+(1+x_1)k_0^3)\phi_{\rho}^a(x_1,b_1)] S_t(x_1) \nonumber \\
	\fl	&\times h_e(x_0,1-x_1,b_0,b_1)\exp[-S_B(t_e^1)-S_{\rho}(t_e^1)] +\alpha_s(t_e^2)r_{\rho1}(E_q-k_0^3)(\phi_{\rho}^v(x_1,b_1) \nonumber\\ \fl&-\phi_{\rho}^a(x_1,b_1)) 
		h_e(1-x_1,x_0,b_1,b_0)S_t(x_0)\exp[-S_B(t_e^2)-S_{\rho}(t_e^2)]\},
\end{eqnarray}
\begin{eqnarray} \label{eq:meN}
	\fl \mathcal{M}_{e,N}=&-16\sqrt{2}\pi^2C_Ff_Bm_B^2 \int k_{0\perp}dk_{0\perp}\int dx_0dx_1dx_2 \int b_0db_0b_2db_2
		\left(\frac{m_B}{2}+\frac{|k_{0\perp}|^2}{2x_0^2m_B}\right) \nonumber\\ \fl&\times K(\vec{k}_0)(E_Q+m_Q)J_0(k_{0\perp}b_0) 
		r_{\rho2}\{\alpha_s(t_{ne}^1) [x_2(E_q-k_0^3)\phi_{\rho}^T(x_1,b_0) \nonumber\\ \fl&\times (\phi_{\rho}^v(x_2,b_2)-\phi_{\rho}^a(x_2,b_2))] 
		\exp[-S_B(t_{ne}^1)-S_{\rho}(t_{ne}^1)|_{b_1\rightarrow b_0}-S_{\rho}(t_{ne}^1)] \nonumber\\ 
	\fl &\times h_{ne}(x_0,x_2,1-x_1,b_0,b_2) 
		+\alpha_s(t_{ne}^2)[(1-x_2)(E_q+k_0^3)\phi_{\rho}^T(x_1,b_0) \nonumber\\ \fl&\times (\phi_{\rho}^v(x_2,b_2)-\phi_{\rho}^a(x_2,b_2))
		-2r_{\rho1}((2-x_1-x_2)E_q-(x_1-x_2)k_0^3) \nonumber\\ \fl&\times (\phi_{\rho}^v(x_1,b_0)\phi_{\rho}^v(x_2,b_2)+\phi_{\rho}^a(x_1,b_0)\phi_{\rho}^a(x_2,b_2))] 
		h_{ne}(x_0,1-x_2,1-x_1,b_0,b_2) \nonumber\\ 
	\fl &\times \exp[-S_B(t_{ne}^2)-S_{\rho}(t_{ne}^2)|_{b_1\rightarrow b_0}-S_{\rho}(t_{ne}^2)]\},
\end{eqnarray}
\begin{eqnarray} \label{eq:merN}
	\fl \mathcal{M}_{e,N}^R=&16\sqrt{2}\pi^2C_Ff_Bm_B^2\int k_{0\perp}dk_{0\perp}\int dx_0dx_1dx_2\int b_0db_0b_2db_2
		\left(\frac{m_B}{2}+\frac{|k_{0\perp}|^2}{2x_0^2m_B}\right) \nonumber\\ \fl&\times K(\vec{k}_0)(E_Q+m_Q)J_0(k_{0\perp}b_0) 
		r_{\rho1}\phi_{\rho}^T(x_2,b_2)\{\alpha_s(t_{ne}^1)[(1-x_1)(E_q+k_0^3) \nonumber\\ \fl&\times (\phi_{\rho}^v(x_1,b_0)+\phi_{\rho}^a(x_1,b_0))] 
		\exp[-S_B(t_{ne}^1)-S_{\rho}(t_{ne}^1)|_{b_1\rightarrow b_0}-S_{\rho}(t_{ne}^1)] \nonumber\\ 
	\fl &\times h_{ne}(x_0,x_2,1-x_1,b_0,b_2) 
		+\alpha_s(t_{ne}^2)[(1-x_1)(E_q+k_0^3)(\phi_{\rho}^v(x_1,b_0)+\phi_{\rho}^a(x_1,b_0))] \nonumber \\
	\fl &\times h_{ne}(x_0,1-x_2,1-x_1,b_0,b_2)\exp[-S_B(t_{ne}^2)-S_{\rho}(t_{ne}^2)|_{b_1\rightarrow b_0}-S_{\rho}(t_{ne}^2)]\},
\end{eqnarray}
\begin{eqnarray} \label{eq:mepN}
	\fl \mathcal{M}_{e,N}^P=&16\sqrt{2}\pi^2C_Ff_Bm_B^2\int k_{0\perp}dk_{0\perp}\int dx_0dx_1dx_2\int b_0db_0b_2db_2
		\left(\frac{m_B}{2}+\frac{|k_{0\perp}|^2}{2x_0^2m_B}\right) \nonumber\\ 
	\fl &\times K(\vec{k}_0)(E_Q+m_Q) J_0(k_{0\perp}b_0) 
		r_{\rho2}\{\alpha_s(t_{ne}^1)[x_2(E_q+k_0^3)\phi_{\rho}^T(x_1,b_0) \nonumber\\ \fl &\times (\phi_{\rho}^v(x_2,b_2)+\phi_{\rho}^a(x_2,b_2))
		-2r_{\rho1}((1-x_1+x_2)E_q+(1-x_1-x_2)k_0^3) \nonumber\\ \fl&\times (\phi_{\rho}^v(x_1,b_0)\phi_{\rho}^v(x_2,b_2)-\phi_{\rho}^a(x_1,b_0)\phi_{\rho}^a(x_2,b_2))]
		h_{ne}(x_0,x_2,1-x_1,b_0,b_2) \nonumber\\
	\fl &\times \exp[-S_B(t_{ne}^1)-S_{\rho}(t_{ne}^1)|_{b_1\rightarrow b_0}-S_{\rho}(t_{ne}^1)]
		+\alpha_s(t_{ne}^2)[(1-x_2)(E_q-k_0^3) \nonumber\\ \fl&\times \phi_{\rho}^T(x_1,b_0)(\phi_{\rho}^v(x_2,b_2)+\phi_{\rho}^a(x_2,b_2))]
		h_{ne}(x_0,1-x_2,1-x_1,b_0,b_2) \nonumber\\ \fl&\times \exp[-S_B(t_{ne}^2)-S_{\rho}(t_{ne}^2)|_{b_1\rightarrow b_0}-S_{\rho}(t_{ne}^2)]\},
\end{eqnarray}
\begin{eqnarray} \label{eq:faN}
	\fl F_{a,N}=&8\sqrt{2}\pi C_Ff_Bm_B^2 \int dx_1dx_2 \int b_1db_1b_2db_22 r_{\rho1}r_{\rho2}\{\alpha_s(t_a^1)[(1+x_2)
		(\phi_{\rho}^v(x_1,b_1)\phi_{\rho}^v(x_2,b_2) \nonumber\\ \fl& +\phi_{\rho}^a(x_1,b_1)\phi_{\rho}^a(x_2,b_2)) 
		-(1-x_2)(\phi_{\rho}^v(x_1,b_1)\phi_{\rho}^a(x_2,b_2)+\phi_{\rho}^a(x_1,b_1)\phi_{\rho}^v(x_2,b_2))] \nonumber \\
	\fl &\times h_a(1-x_1,x_2,b_1,b_2)S_t(x_2) \exp[-S_{\rho}(t_a^1)-S_{\rho}(t_a^1)] 
		+\alpha_s(t_a^2)[-(2-x_1) \nonumber\\ \fl&\times (\phi_{\rho}^v(x_1,b_1)\phi_{\rho}^v(x_2,b_2)+\phi_{\rho}^a(x_1,b_1)\phi_{\rho}^a(x_2,b_2))
		-x_1(\phi_{\rho}^v(x_1,b_1)\phi_{\rho}^a(x_2,b_2) \nonumber\\ \fl& +\phi_{\rho}^a(x_1,b_1)\phi_{\rho}^v(x_2,b_2))]
		h_a(x_2,1-x_1,b_2,b_1)S_t(x_1) \exp[-S_{\rho}(t_a^2)-S_{\rho}(t_a^2)]\},
\end{eqnarray}
\begin{eqnarray} \label{eq:fapN}
	\fl F_{a,N}^P=&8\sqrt{2}\pi C_Ff_Bm_B^2\chi_B\int dx_1dx_2\int b_1db_1b_2db_2 (-4)\{\alpha_s(t_a^1)[r_{\rho1}(\phi_{\rho}^v(x_1,b_1)-\phi_{\rho}^a(x_1,b_1)) \nonumber\\ \fl&\times \phi_{\rho}^T(x_2,b_2)]
		h_a(1-x_1,x_2,b_1,b_2)S_t(x_2)\exp[-S_{\rho}(t_a^1)-S_{\rho}(t_a^1)] \nonumber \\
	\fl	&+\alpha_s(t_a^2) [r_{\rho2}\phi_{\rho}^T(x_1,b_1)(\phi_{\rho}^v(x_2,b_2)+\phi_{\rho}^a(x_2,b_2))] 
		h_a(x_2,1-x_1,b_2,b_1)S_t(x_1) \nonumber\\ 
	\fl &\times \exp[-S_{\rho}(t_a^2)-S_{\rho}(t_a^2)]\},
\end{eqnarray}
\begin{eqnarray} \label{eq:maN}
	\fl \mathcal{M}_{a,N}=&16\sqrt{2}\pi^2C_Ff_Bm_B^2\int k_{0\perp}dk_{0\perp}\int dx_0dx_1dx_2\int b_0db_0b_1db_1
		\left(\frac{m_B}{2}+\frac{|k_{0\perp}|^2}{2x_0^2m_B}\right) \nonumber\\ \fl&\times K(\vec{k}_0)(E_Q+m_Q) J_0(k_{0\perp}b_0) 
		\{\alpha_s(t_{na}^1)[((-r_{\rho1}^2x_1+r_{\rho2}^2(x_2-1))E_q \nonumber\\ \fl& +(r_{\rho1}^2x_1+r_{\rho2}^2(x_2-1))k_0^3)\phi_{\rho}^T(x_1,b_1)\phi_{\rho}^T(x_2,b_1)
		+2r_{\rho1}r_{\rho2}E_q(\phi_{\rho}^v(x_1,b_1)\phi_{\rho}^v(x_2,b_1) \nonumber\\ \fl& +\phi_{\rho}^a(x_1,b_1)\phi_{\rho}^a(x_2,b_1))
		-2r_{\rho1}r_{\rho2}k_0^3 (\phi_{\rho}^v(x_1,b_1)\phi_{\rho}^a(x_2,b_1)+\phi_{\rho}^a(x_1,b_1)\phi_{\rho}^v(x_2,b_1))] \nonumber \\
	\fl	&\times h_{na}^1(1-x_1,x_2,b_0,b_1)\exp[-S_B(t_{na}^1)-S_{\rho}(t_{na}^1)-S_{\rho}(t_{na}^1)|_{b_2\rightarrow b_1}] 
		+\alpha_s(t_{na}^2) \nonumber\\ \fl&\times[((r_{\rho1}^2(x_1-1)-r_{\rho2}^2x_2)E_q+(r_{\rho1}^2(1-x_1)-r_{\rho2}^2x_2)k_0^3)\phi_{\rho}^T(x_1,b_1)\phi_{\rho}^T(x_2,b_1)] \nonumber\\ 
	\fl&\times \exp[-S_B(t_{na}^2)-S_{\rho}(t_{na}^2)-S_{\rho}(t_{na}^2)|_{b_2\rightarrow b_1}] 
		h_{na}^2(1-x_1,x_2,b_0,b_1)\},
\end{eqnarray}
\begin{eqnarray} \label{eq:marN}
	\fl \mathcal{M}_{a,N}^R=&16\sqrt{2}\pi^2C_Ff_Bm_B^2\int k_{0\perp}dk_{0\perp}\int dx_0dx_1dx_2\int b_0db_0b_1db_1
		\left(\frac{m_B}{2}+\frac{|k_{0\perp}|^2}{2x_0^2m_B}\right) \nonumber\\ \fl&\times K(\vec{k}_0)(E_Q+m_Q) J_0(k_{0\perp}b_0) 
		\{\alpha_s(t_{na}^1)[r_{\rho2}((2-x_2)E_q+x_2k_0^3)\phi_{\rho}^T(x_1,b_1) \nonumber\\ \fl&\times (\phi_{\rho}^v(x_2,b_1)+\phi_{\rho}^a(x_2,b_1))
		- r_{\rho1}((1+x_1)E_q-(1-x_1)k_0^3) \nonumber\\ \fl&\times (\phi_{\rho}^v(x_1,b_1)-\phi_{\rho}^a(x_1,b_1))\phi_{\rho}^T(x_2,b_1)]
		\exp[-S_B(t_{na}^1)-S_{\rho}(t_{na}^1)-S_{\rho}(t_{na}^1)|_{b_2\rightarrow b_1}] \nonumber \\
	\fl &\times h_{na}^1(1-x_1,x_2,b_0,b_1)
		+\alpha_s(t_{na}^2)[r_{\rho2}x_2(E_q+k_0^3)\phi_{\rho}^T(x_1,b_1)(\phi_{\rho}^v(x_2,b_1) \nonumber\\ \fl& + \phi_{\rho}^a(x_2,b_1))
		-r_{\rho1}(1-x_1)(E_q-k_0^3)(\phi_{\rho}^v(x_1,b_1)-\phi_{\rho}^a(x_1,b_1)) \phi_{\rho}^T(x_2,b_1)] \nonumber\\
	\fl &\times h_{na}^2(1-x_2,x_3,b_1,b_2)\exp[-S_B(t_{na}^2)-S_{\rho}(t_{na}^2)-S_{\rho}(t_{na}^2)|_{b_2\rightarrow b_1}]\},
\end{eqnarray}
\begin{eqnarray} \label{eq:mapN}
	\fl \mathcal{M}_{a,N}^P=&16\sqrt{2}\pi^2C_Ff_Bm_B^2\int k_{0\perp}dk_{0\perp}\int dx_0dx_1dx_2\int b_0db_0b_1db_1
		\left(\frac{m_B}{2}+\frac{|k_{0\perp}|^2}{2x_0^2m_B}\right) \nonumber\\ \fl&\times K(\vec{k}_0)(E_Q+m_Q) J_0(k_{0\perp}b_0) 
		\{\alpha_s(t_{na}^1)[((-r_{\rho1}^2x_1+r_{\rho2}^2(x_2-1))E_q \nonumber\\ \fl& +(r_{\rho1}^2x_1+r_{\rho2}^2(x_2-1))k_0^3)\phi_{\rho}^T(x_1,b_1)\phi_{\rho}^T(x_2,b_1)
		+2r_{\rho1}r_{\rho2}E_q(\phi_{\rho}^v(x_1,b_1)\phi_{\rho}^v(x_2,b_1) \nonumber\\ \fl& +\phi_{\rho}^a(x_1,b_1)\phi_{\rho}^a(x_2,b_1))
		-2r_{\rho1}r_{\rho2}k_0^3(\phi_{\rho}^v(x_1,b_1)\phi_{\rho}^a(x_2,b_1)+\phi_{\rho}^a(x_1,b_1)\phi_{\rho}^v(x_2,b_1))] \nonumber \\
	\fl &\times h_{na}^1(1-x_1,x_2,b_0,b_1)\exp[-S_B(t_{na}^1)-S_{\rho}(t_{na}^1)-S_{\rho}(t_{na}^1)|_{b_2\rightarrow b_1}] 
		+\alpha_s(t_{na}^2) \nonumber\\ \fl&\times [((r_{\rho1}^2(x_1-1)-r_{\rho2}^2x_2)E_q+(r_{\rho1}^2(1-x_1)-r_{\rho2}^2x_2)k_0^3)\phi_{\rho}^T(x_1,b_1)\phi_{\rho}^T(x_2,b_1)] \nonumber\\
	\fl &\times \exp[-S_B(t_{na}^2)-S_{\rho}(t_{na}^2)-S_{\rho}(t_{na}^2)|_{b_2\rightarrow b_1}] h_{na}^2(1-x_1,x_2,b_0,b_1)\}.
\end{eqnarray}

For the transverse amplitude, we define the substitution
\begin{eqnarray} \label{subst}
	S_1 = \left\{\phi_{\rho}^{v(a)}(x_1)\to\phi_{\rho}^{a(v)}(x_1),\ r_{\rho1}\to -r_{\rho1}\right\},
\end{eqnarray}
and the results can be derived from the parallel case by the application of the substitution $S_1$:
\begin{eqnarray}
	F_{e,T} &= -i\left. F_{e,N} \right|_{S_1},\ &F_{a,T}^{(P)} = -i\left. F_{a,N}^{(P)} \right|_{S_1}, \nonumber \\
	\mathcal{M}_{e,T}^{(P)} &= -i\left. \mathcal{M}_{e,N}^{(P)} \right|_{S_1},\ &\mathcal{M}_{e,T}^R = i\left. \mathcal{M}_{e,N}^R \right|_{S_1}, \nonumber \\
	\mathcal{M}_{a,T}^{(R)} &= -i\left. \mathcal{M}_{a,N}^{(R)} \right|_{S_1},\ &\mathcal{M}_{a,T}^P = i\left. \mathcal{M}_{a,N}^P \right|_{S_1}.
\end{eqnarray}
Furthermore, by the analysis of the Lorentz structure, there are some relations to connect the residual amplitudes with those given above
\begin{eqnarray} \label{eq:ferpfarLNT}
    \fl F_{e,L(N,T)}^R&=F_{e,L(N,T)}, \quad F_{e,L(N,T)}^P=0, \quad F_{a,L(N)}^R &= F_{a,L(N)}, \quad F_{a,T}^R = -F_{a,T}.
\end{eqnarray}

Regarding the results summarized in Eqs.~(\ref{eq:feL})-(\ref{eq:ferpfarLNT}), we should provide some clarifications. 
In the parallel amplitudes listed in Eqs.~(\ref{eq:feN})-(\ref{eq:mapN}), we employ an extra subscript $1$ or $2$ to indicate the particular numbered $\rho$ meson from which the $r_\rho$ comes. 
This notation induces no numerical complexity while allows us to express the transverse amplitudes through a straightforward substitution $S_1$. 
The integral ranges of various variables are written explicitly in Eq.~(\ref{eq:feL}), with $x_0^{u,d}=1/2\pm\sqrt{1/4-|k_{0\perp}|^2/m_B^2}$, and the relevant integral ranges in other equations remain the same. 
The hard functions $h$ derived from the Fourier transformation are defined by 
\begin{eqnarray}
	\fl &\ h_e(x_1,x_2,b_1,b_2)=&K_0(\sqrt{x_1x_2}m_Bb_1) \nonumber\\ 
	\fl &&\times [\theta(b_1-b_2)K_0(\sqrt{x_2}m_Bb_1)I_0(\sqrt{x_2}m_Bb_2) + (b_1 \leftrightarrow b_2)], \nonumber \\
	\fl &\ h_a(x_1,x_2,b_1,b_2)=&K_0(-i\sqrt{x_1x_2}m_Bb_1) \nonumber\\ 
	\fl &&\times [\theta(b_1-b_2)K_0(-i\sqrt{x_2}m_Bb_1)I_0(-i\sqrt{x_2}m_Bb_2)+ (b_1 \leftrightarrow b_2)], \nonumber \\
	\fl h_{ne}(&x_1,x_2,x_3,b_1,b_2)\ =&K_0(-i\sqrt{x_2x_3}m_Bb_2) \nonumber\\
	\fl &&\times [\theta(b_1-b_2)K_0(\sqrt{x_1x_3}m_Bb_1)I_0(\sqrt{x_1x_3}m_Bb_2)+ (b_1 \leftrightarrow b_2)], \nonumber \\
	\fl &h_{na}^1(x_1,x_2,b_1,b_2)=&K_0(-i\sqrt{x_1x_2}m_Bb_1) \nonumber \\
	\fl	&&\times[\theta(b_1-b_2) K_0(-i\sqrt{x_1x_2}m_Bb_1)I_0(-i\sqrt{x_1x_2}m_Bb_2)+ (b_1 \leftrightarrow b_2)], \nonumber \\
	\fl &h_{na}^2(x_1,x_2,b_1,b_2)=&K_0(\sqrt{x_1+x_2-x_1x_2}m_Bb_1) [\theta(b_1-b_2) \nonumber \\
	\fl	&&\times K_0(-i\sqrt{x_1x_2}m_Bb_1)I_0(-i\sqrt{x_1x_2}m_Bb_2)+ (b_1 \leftrightarrow b_2)],
\end{eqnarray}
where $\theta$ is the Heaviside step function, $K_0$, $I_0$, and $J_0$ in Eqs.~(\ref{eq:feL})-(\ref{eq:ferpfarLNT}) are different types of Bessel functions. 
The renormalization scales $t$ are determined as the maximum virtuality in each diagram to suppress the impact of high order corrections. 
Their values are represented by 
\begin{eqnarray}
	t_e^1=&\max(\sqrt{1-x_1}m_B, 1/b_0, 1/b_1), \nonumber \\
    t_e^2=&\max(\sqrt{x_0}m_B, 1/b_0, 1/b_1), \nonumber \\
	t_a^1=&\max(\sqrt{x_2}m_B, 1/b_1, 1/b_2), \nonumber \\
	t_a^2=&\max(\sqrt{1-x_1}m_B, 1/b_1, 1/b_2), \nonumber \\
	t_{ne}^1=&\max(\sqrt{x_0(1-x_1)}m_B, \sqrt{x_2(1-x_1)}m_B, 1/b_0, 1/b_2), \nonumber \\
    t_{ne}^2=&\max(\sqrt{x_0(1-x_1)}m_B, \sqrt{(1-x_2)(1-x_1)}m_B, 1/b_0, 1/b_2), \nonumber \\
	t_{na}^1=&\max(\sqrt{(1-x_1)x_2}m_B, 1/b_0, 1/b_1), \nonumber \\
    t_{na}^2=&\max(\sqrt{(1-x_1)x_2}m_B, \sqrt{1-x_1+x_1x_2}m_B, 1/b_0, 1/b_1).
\end{eqnarray}
As discussed in Sec.~\ref{subsec:frame}, the overlap of the collinear and soft enhancements gives rise to the double logarithm $\alpha_s\ln^2k_T$, which should be treated by resummation technique and generate the Sudakov factor \cite{Li:1992nu, Li:1992ce, Li:1994cka, Botts:1989kf}. 
The exponents for Sudakov and single ultraviolet logarithms associated with $B$ and $\rho$ mesons are given by 
\begin{eqnarray} \label{eq:sudakovBrho}
	S_B(t)=&s(x_0,b_0,m_B)-\frac{1}{\beta_1} \ln\frac{\ln(t/\Lambda_{\mathrm{QCD}})}{\ln(1/(b_0\Lambda_{\mathrm{QCD}}))}, \nonumber \\
	S_\rho(t)=&s(x_1,b_1,m_B)+s(1-x_1,b_1,m_B) -\frac{1}{\beta_1}\ln\frac{\ln(t/\Lambda_{\mathrm{QCD}})}{\ln(1/(b_1\Lambda_{\mathrm{QCD}}))}.
\end{eqnarray}
The Sudakov factor $s(x, b, Q)$ is defined as follows
\begin{eqnarray} \label{eq:sudakov}
    \fl s(x,b,Q) =&\frac{A^{(1)}}{2\beta_1}\hat{q}\ln\left(\frac{\hat{q}}{\hat{b}}\right)-\frac{A^{(1)}}{2\beta_1}\left(\hat{q}-\hat{b}\right)
		+\frac{A^{(2)}}{4\beta_1^2}\left(\frac{\hat{q}}{\hat{b}}-1\right) 
		-\left[\frac{A^{(2)}}{4\beta_1^2}-\frac{A^{(1)}}{4\beta_1}\ln\left(\frac{e^{2\gamma_E-1}}{2}\right)\right] \nonumber\\ \fl&\times \ln\left(\frac{\hat{q}}{\hat{b}}\right) 
		+\frac{A^{(1)}\beta_2}{4\beta_1^3}\hat{q}\left[\frac{\ln(2\hat{q})+1}{\hat{q}}-\frac{\ln(2\hat{b})+1}{\hat{b}}\right] 
		+\frac{A^{(1)}\beta_2}{8\beta_1^3}\left[\ln^2(2\hat{q})-\ln^2(2\hat{b})\right] \nonumber \\
	\fl &+\frac{A^{(1)}\beta_2}{8\beta_1^3}\ln\left(\frac{e^{2\gamma_E-1}}{2}\right)\left[\frac{\ln(2\hat{q})+1}{\hat{q}}-\frac{\ln(2\hat{b})+1}{\hat{b}}\right] \nonumber \\
	\fl &-\frac{A^{(2)}\beta_2}{16\beta_1^4}\left[\frac{2\ln(2\hat{q})+3}{\hat{q}}-\frac{2\ln(2\hat{b})+3}{\hat{b}}\right] 
		-\frac{A^{(2)}\beta_2}{16\beta_1^4}\frac{\hat{q}-\hat{b}}{\hat{b}^2}\left[2\ln(2\hat{b})+1\right] \nonumber \\
	\fl &+\frac{A^{(2)}\beta_2^2}{1728\beta_1^6}\left[\frac{18\ln^2(2\hat{q})+30\ln(2\hat{q})+19}{\hat{q}^2} 
		-\frac{18\ln^2(2\hat{b})+30\ln(2\hat{b})+19}{\hat{b}^2}\right] \nonumber \\
	\fl &+\frac{A^{(2)}\beta_2^2}{432\beta_1^6}\frac{\hat{q}-\hat{b}}{\hat{b}^3}\left[9\ln^2(2\hat{b})+6\ln(2\hat{b})+2\right],
\end{eqnarray}
with the abbreviations
\begin{eqnarray} \label{eq:qphat}
  \hat{q}\equiv \ln\left(xQ/\left(\sqrt{2}\Lambda_{\textup{QCD}}\right)\right),\quad
  \hat{b}\equiv \ln\left(1/\left(b\Lambda_{\textup{QCD}}\right)\right),
\end{eqnarray}
and the coefficients
\begin{eqnarray} \label{eq:betaandA}
	\beta_1 &= \frac{33-2n_f}{12}, \ \beta_2=\frac{153-19n_f}{24}, \ A^{(1)} = \frac{4}{3}, \nonumber \\
	A^{(2)} &= \frac{67}{9}-\frac{\pi^2}{3} -\frac{10}{27}n_f+\frac{8}{3}\beta_1\ln\left(\frac{e^{\gamma_E}}{2}\right).
\end{eqnarray}
The resummation of the double logarithm $\alpha_s\ln^2x$ gives rise to the threshold factor $S_t(x)$, which suppresses the end-point contribution and improves the PQCD calculation
\cite{Li:2001ay, Li:2002mi, Kurimoto:2001zj}.
Its expression reads
\begin{eqnarray} \label{eq:st}
  S_t(x)=\frac{2^{1+2c}\Gamma(3/2+c)}{\sqrt{\pi}\Gamma(1+c)}[x(1-x)]^c,
\end{eqnarray}
with $c=0.3$.

\section{$\rho$ meson distribution amplitudes \label{app:rholcda}}
The dependence of the LCDAs $\phi$ on the transverse momentum $k_{q\perp}$ of the constituent quark is postulated to be Gaussian distribution and can be factorized out with the $x$ part.
This assumption has been thoroughly investigated in Ref.~\cite{Wei:2002iu}.
With the Fourier transformation from transverse momentum $k_{q\perp}$ space to transverse coordinate $b$ space, we can express the LCDAs $\phi$ in a simple form
\begin{eqnarray}
	\phi(x,b)=\phi(x)\exp\left(-b^2/(4\beta^2)\right),
\end{eqnarray}
where $\phi(x)$ represents the LCDA with respect to longitudinal distribution, and the oscillation parameter is taken to be $\beta = 4.0~\textrm{GeV}^{-1}$~\cite{Wei:2002iu}.

By the conventional conformal expansion \cite{Braun:2003rp}, the twist-2 and twist-3 LCDAs of the $\rho$ meson can be expressed in terms of the Gegenbauer polynomials $C_n^{j/2}$.
For the $\rho$ meson with the longitudinal polarization vector, the LCDAs read \cite{Ball:2004rg, Ball:2007rt}
\begin{eqnarray} \label{eq:phiVVs}
	\fl \phi_\rho(x) =& \frac{f_\rho^\parallel}{2\sqrt{2N_c}} 6x\bar{x}\left[1+a_{1\rho}^\parallel C_1^{3/2}(t) + a_{2\rho}^\parallel C_2^{3/2}(t)\right], \quad
	\phi_\rho^s(x) = \frac{f_\rho^\perp}{2\sqrt{2N_c}} \frac{1}{2} \frac{d}{dx} \phi_\rho^u(x),
\end{eqnarray}
\begin{eqnarray} \label{eq:phiVt}
	\fl \phi_\rho^t(x) =& \frac{f_\rho^\perp}{2\sqrt{2N_c}}\Bigl\{3t C_1^{1/2}(t) + 3t a_{1\rho}^\perp C_2^{1/2}(t)
		+ \left(3t a_{2\rho}^\perp + 15\kappa_{3\rho}^\perp - 3\lambda_{3\rho}^\perp/2\right) C_3^{1/2}(t) \nonumber\\ \fl& + 5\omega_{3\rho}^\perp C_4^{1/2}(t)
		+\frac{3}{2}\frac{m_{q1}+m_{q2}}{m_\rho}\frac{f_\rho^\parallel}{f_\rho^\perp}\left[1+8ta_{1\rho}^\parallel + 3(7-30x\bar{x})a_{2\rho}^\parallel \right.\nonumber \\
	\fl &\left. +t\left(1+3a_{1\rho}^\parallel+ 6a_{2\rho}^\parallel\right)\ln\bar{x} - t\left(1-3a_{1\rho}^\parallel+6a_{2\rho}^\parallel\right)\ln x\right] \Bigr\}, 
\end{eqnarray}
\begin{eqnarray} \label{eq:phiVu}
	\fl \phi_\rho^u(x) =& 6x\bar{x}\left[1+\left(a_{1\rho}^\perp/3 + 5\kappa_{3\rho}^\perp/3\right) C_1^{3/2}(t) + \left(a_{2\rho}^\perp/6 + 5\omega_{3\rho}^\perp/18\right)C_{2}^{3/2}(t) \right.\nonumber \\
	\fl &\left. - \left(\lambda_{3\rho}^\perp/20\right) C_3^{3/2}(t)\right]
		+3\frac{m_{q1}+m_{q2}}{m_\rho}\frac{f_\rho^\parallel}{f_\rho^\perp}\left[x\bar{x}\left(1+2ta_{1\rho}^\parallel+3\left(7-5x\bar{x}\right)a_{2\rho}^\parallel\right) \right.\nonumber\\
	\fl &\left.+ \left(1+3a_{1\rho}^\parallel+6a_{2\rho}^\parallel\right)\bar{x}\ln\bar{x}  
		+ \left(1-3a_{1\rho}^\parallel+6a_{2\rho}^\parallel\right) x\ln x \right],
\end{eqnarray}
while for the transversely polarized states, the LCDAs are described by
\begin{eqnarray} \label{eq:phiVTVa}
	\fl \phi_\rho^T(x) =& \frac{f_\rho^\perp}{2\sqrt{2N_c}} 6x\bar{x} \left[1+a_{1\rho}^\perp C_1^{3/2}(t) + a_{2\rho}^\perp C_2^{3/2}(t)\right], \quad
	\phi_\rho^a(x) = \frac{f_\rho^\parallel}{2\sqrt{2N_c}}\frac{1}{4}\frac{d}{dx}\phi_\rho^o(x),
\end{eqnarray}
\begin{eqnarray} \label{eq:phiVv}
	\fl \phi_\rho^v(x) =& \frac{f_\rho^\parallel}{2\sqrt{2N_c}} \Big\{3(1+t^2)/4 + \left(3/2\right)t^2a_{1\rho}^\parallel C_1^{1/2}(t)
		+ \left(3a_{2\rho}^\parallel/7 + 5\zeta_{3\rho}^\parallel\right) 2C_2^{1/2}(t) \nonumber \\
	\fl & + \left(5\kappa_{3\rho}^\parallel - 15\lambda_{3\rho}^\parallel/16 + 15\tilde{\lambda}_{3\rho}^\parallel/8 \right) 2C_3^{1/2}(t)
		+ \left(9a_{2\rho}^\parallel/14 + 15\omega_{3\rho}^\parallel/4 \right.\nonumber\\ \fl&\left. - 15\tilde{\omega}_{3\rho}^\parallel/8 \right) C_4^{1/2}(t) 
		+ \frac{3}{2}\frac{m_{q1}+m_{q2}}{m_\rho}\frac{f_\rho^\perp}{f_\rho^\parallel} \Bigl[2 + 9ta_{1\rho}^\perp + 2\left(11 - 30x\bar{x}\right)a_{2\rho}^\perp \nonumber\\
	\fl	&+ \left(1 - 3a_{1\rho}^\perp + 6a_{2\rho}^\perp\right) \ln x + \left(1 + 3a_{1\rho}^\perp + 6a_{2\rho}^\perp\right) \ln\bar{x} \Bigr] \Big\}, 
\end{eqnarray}
\begin{eqnarray} \label{eq:phiVo}
	\fl \phi_\rho^o(x) =& 6x\bar{x}\Bigl[1 + \left(a_{1\rho}^\parallel/3 + 20\kappa_{3\rho}^\parallel/9\right) C_1^{3/2}(t)
		+ \left(\tilde{\lambda}_{3\rho}^\parallel/4 - \lambda_{3\rho}^\parallel/8\right) C_3^{3/2}(t) \nonumber \\
	\fl &+ \left(a_{2\rho}^\parallel/6 + 10\zeta_{3\rho}^\parallel/9 + 5\omega_{3\rho}^\parallel/12 - 5\tilde{\omega}_{3\rho}^\parallel/24\right) C_2^{3/2}(t)\Bigr] 
		+ 6\frac{m_{q1}+m_{q2}}{m_\rho}\frac{f_\rho^\perp}{f_\rho^\parallel} \nonumber\\ \fl&\times \left\{x\bar{x}\left[2 + 3ta_{1\rho}^\perp + 2\left(11 - 10x\bar{x}\right)a_{2\rho}^\perp\right] +\bar{x}\ln\bar{x}\left(1 + 3a_{1\rho}^\perp + 6a_{2\rho}^\perp\right)\right. \nonumber \\
	\fl &\left. + x\ln x\left(1 - 3a_{1\rho}^\perp + 6a_{2\rho}^\perp\right) \right\},
\end{eqnarray}
where $t = 2x-1$ and $\bar{x}=1-x$. The mass of the constituent quark (or antiquark) is denoted by $m_{q1}$ (or $m_{q2}$). We neglect the mass difference between the light ($u$ and $d$) quarks, and consequently, the terms proportional to $m_{q1}-m_{q2}$ in the LCDAs are omitted.

The hadronic parameters estimated from the QCD sum rules at the scale of $1~\textrm{GeV}$ are presented as follows \cite{Ball:2007rt}
\begin{eqnarray}
	\fl a_{1\rho}^{\parallel, \perp} &= \kappa_{3\rho}^{\parallel,\perp} = \lambda_{3\rho}^{\parallel, \perp} = \tilde{\lambda}_{3\rho}^\parallel = 0, \quad
		a_{2\rho}^\parallel = 0.15\pm 0.07, \quad a_{2\rho}^\perp = 0.14\pm 0.06, \nonumber \\
	\fl \omega_{3\rho}^\parallel &= 0.15\pm 0.05, \ \omega_{3\rho}^\perp = 0.55\pm 0.25, \ 
		\tilde{\omega}_{3\rho}^\parallel = -0.09\pm 0.03, \ \zeta_{3\rho}^\parallel = 0.030\pm 0.010.
\end{eqnarray}

\section*{References}

\end{document}